\documentclass[12pt]{iopart}

\usepackage{iopams}  
\usepackage{citesort}  
\usepackage[dvips]{epsfig}                   

\newcommand{\beqa}{\begin{eqnarray}}
\newcommand{\eeqa}{\end{eqnarray}}
\newcommand{\beq}{\begin{equation}}
\newcommand{\eeq}{\end{equation}}

\newcommand{\Dslash}{D\hspace{-.5em}/\hspace{.15em}}
\newcommand{\qslash}{q\hspace{-.5em}/\hspace{.15em}}

\newcommand{\pslash}{p\hspace{-.5em}/\hspace{.15em}}
\newcommand{\Pslash}{P\hspace{-.5em}/\hspace{.15em}}

\eqnobysec

\begin{document}

\topical[Infrared Properties of QCD]{Infrared Properties of QCD from Dyson-Schwinger equations}
\author{Christian S. Fischer}

\address{GSI, Planckstr. 1, 64291 Darmstadt, Germany}
\ead{christian.fischer@physik.tu-darmstadt.de}
\begin{abstract}	
I review recent results on the infrared properties of QCD from Dyson-Schwinger equations. 
The topics include infrared exponents of one-particle irreducible Green's functions, the 
fixed point behaviour of the running coupling at zero momentum, the pattern of dynamical 
quark mass generation and properties of light mesons.
\end{abstract}


\tableofcontents

\markboth{Infrared Properties of QCD from Dyson-Schwinger equations}{Infrared Properties of QCD from Dyson-Schwinger equations}

\section{Introduction}

Our understanding how observable properties of hadrons emerge from the 
underlying structure of the strong interaction is still far from complete. 
Physical phenomena encountered at large momentum transfers are very well 
described by perturbation theory. {\it Asymptotic freedom} allows high 
energy probes to picture hadrons as lumps of weakly interacting quarks 
and gluons. This picture, however, starts to break down at intermediate 
momenta and is surely inadequate at energies below a view hundred MeV. At 
such scales the interaction is strong enough to invalidate perturbation 
theory and one has to employ completely different methods to deal with 
what can be called {\it Strong QCD}. 

There are two fundamental low energy properties of QCD: confinement and dynamical 
chiral symmetry breaking. Both are entirely strong coupling effects in the 
sense that they cannot be accounted for at any order in perturbation theory. 
They are presumably related to each other, although the detailed structure of 
this relation is not yet clear. Both together are responsible for the complexity 
of the experimental hadron spectrum. The internal structure of hadrons 
beyond the simple valence quark model is one of the central issues in 
contemporary low energy hadron physics. Nonperturbative methods such as lattice 
Monte-Carlo simulations \cite{Greensite:2003bk,Chandrasekharan:2004cn}, the 
exact renormalisation group (see \cite{Pawlowski:2005xe} and references therein) or 
the Green's function approach employing Dyson-Schwinger and Bethe-Salpeter 
equations \cite{Roberts:1994dr,Alkofer:2000wg,Roberts:2000aa,Maris:2003vk} 
are appropriate tools to explore the details of these structures.

Dyson-Schwinger and Bethe-Salpeter equations (DSEs/BSEs) are the equations of 
motion of the Green's functions of a field theory. In QCD the study of the 
nonperturbative behaviour of these functions is interesting for several reasons. 
Confinement mechanisms like the Kugo-Ojima criterion or the Gribov-Zwanziger 
scenario are related to the infrared behaviour of the ghost and gluon propagators. 
Nonperturbative definitions of the running coupling can be given in terms of 
renormalisation group invariants constructed from two and three point functions. 
Effects of dynamical chiral symmetry breaking are encoded in the quark propagator 
and the quark-gluon interaction. Furthermore, properties of bound states and 
resonances can be determined from the n-point functions of the theory. 

The machinery of DSEs and BSEs has been studied extensively in the last years.
From a technical point of view these continuum Green's functions methods and lattice 
Monte-Carlo simulations are complementary to each other. Lattice calculations 
contain all effects from quantum fluctuations and are therefore the only 
{\it ab initio} approach available so far. However, lattice results are 
limited to a comparably small momentum range and potentially suffer from finite 
volume effects in the infrared. The implementation of realistic light quark 
masses still awaits faster CPUs and better algorithms.  

Dyson-Schwinger equations, on the other hand, can be solved analytically in 
the infrared and provide numerical solutions for a large momentum range. All 
aspects of chiral symmetry are respected such that the properties and effects 
of light quarks can be determined systematically with reasonable effort. 
A technical challenge of DSEs is that they form an infinite tower of equations 
which are coupled to each other. Thus, in order to obtain a closed system of equations
for n-point functions one has to introduce approximations for some m-point 
functions with $m > n$ (unless one works in certain kinematical limits, see
subsection \ref{IR}). These truncations have to be controlled, {\it e.g.}
by comparison to corresponding lattice calculations in the momentum region where
lattice results are available. In turn, however, the Dyson-Schwinger approach provides 
important information in momentum and quark mass regions that are not (yet) 
accessible on the lattice.

A note on gauge independence is in order here. In general, Green's functions are gauge dependent objects. 
Certainly, confinement and dynamical mass generation are experimentally observable phenomena 
and as such must have gauge independent theoretical signatures. Confinement is encoded in the 
long distance behaviour of the (gauge invariant) Wilson loop and the strength of dynamical 
chiral symmetry breaking is given by its (gauge invariant) order parameter, the chiral 
condensate. On the other hand it is very well possible that the detailed mechanism that 
generates these quantities depends on the gauge. Furthermore possible order parameters separating 
the confining from the deconfining phase of gauge theories may only be identifiable after gauge 
fixing. Therefore the approach to study the theoretical 
structures leading to these phenomena in different gauge fixed formulations of QCD is 
justified, interesting and well established (see {\it e.g.} \cite{Greensite:2003bk} and 
reference therein).

This review is intended as an overview on selected topics of nonperturbative QCD. I 
concentrate on some questions of low energy QCD which to my opinion can be answered 
particularly well in the DSE/BSE approach. Owing to limited space I focus on QCD 
at zero temperature and zero density and discuss in some detail recent results that have 
not yet been included in other reviews. Interesting developments within the DSE/BSE approach
at finite temperature or density are described in \cite{Roberts:2000aa,Maas:2005ym,Maas:2005hs,Nickel:2006vf}.  
Furthermore I can only give a brief account on results for various meson observables. 
Reviews on this subject can be found in \cite{Alkofer:2000wg,Maris:2003vk,Maris:2005tt}. 
Studies of the baryon sector of QCD employing Faddeev equations are still exploratory. 
Recent results can be found 
in \cite{Bloch:1999rm,Ahlig:2000qu,Oettel:2000jj,Hecht:2002ej,Alkofer:2004yf}.

\section{The Yang-Mills sector of QCD \label{YM}}

Confinement, {\it i.e.} the absence of (coloured) quarks and gluons from the observable spectrum
is widely believed to be generated in the gauge sector of 
QCD\footnote{A different view has been advocated by Gribov, for a review of his ideas 
see \cite{Dokshitzer:2004ie}.}. This phenomenon should be reflected  
in the infrared properties of the dressed one-particle irreducible Green's functions 
of Yang-Mills theory. This is one of the reasons why these functions have been studied
in the past. Another reason is that elements of Yang-Mills theory, as {\it e.g.} the 
dressed gluon propagator, are vital ingredients for a description of hadrons via 
Bethe--Salpeter equations \cite{Alkofer:2000wg,Roberts:2000aa}\footnote{I discuss
this in more detail in section \ref{mesons}.}.   
Older works on this subject are based on the idea of infrared slavery and assume a gluon 
propagator that is strongly singular at zero momentum. In the past years this picture 
has changed and recent studies employing Dyson--Schwinger equations
\cite{vonSmekal:1997is,Atkinson:1998tu,Watson:2001yv,Zwanziger:2001kw,Lerche:2002ep,Zwanziger:2002ia}
or lattice Monte-Carlo simulations
\cite{Bonnet:2000kw,Bonnet:2001uh,Langfeld:2001cz,Cucchieri:2000kw}
indicate quite the opposite: an infrared finite or even vanishing gluon propagator. 
Nonperturbative infrared singularities, however, are predicted to occur in the ghost 
propagator, the vertices and also in most of the other n-point functions of Yang-Mills theory.
I explain the logical structure of the arguments leading to this result in subsection \ref{IR}
and discuss some interesting consequences in the remaining parts of this section.
In order to appreciate these results it is useful to first recall some aspects of confinement 
related to Green's functions.

\subsection{Aspects of confinement \label{conf}}
 
 QCD is the quantum field theory of quarks and gluons. In Euclidean space-time the partition 
 function is given by\footnote{Detailed explanations
 of the physical content of the QCD partition function is given in many textbooks, see {\it e.g.}
 \cite{Halzen:1984mc,Peskin:1995ev}. An introduction into path integral methods in quantum field theories 
 is given {\it e.g.} in \cite{Rivers:1987hi}.} 
\beqa
\hspace*{-2cm}
Z[J,\eta,\bar{\eta}] &=& 
\int {\cal D} [A \bar{\Psi} \Psi] \exp\left\{-\int d^4x \left(\bar{\Psi} \left( -\Dslash + m\right) \Psi + 
\frac{1}{4} F_{\mu \nu}^a F_{\mu \nu}^a\right) \right.\nonumber\\
&&\hspace*{3cm} \left.+ \int d^4x 
\left(A^a_\mu J^a_\mu+\bar{\eta}\Psi+\bar{\Psi}\eta \right)\right\}. 
\label{genfunc0}
\eeqa 
Quarks are represented by the Dirac fields $\Psi$ and $\bar{\Psi}$.
Local gauge symmetry of the quark fields demands the introduction of a vector field $A_\mu^a$,
which represents gluons. External sources for these fields are denoted by 
$\bar{\eta}$, $\eta$ and $J^a_\mu$. The gluon field strength $F_{\mu \nu}^a$ is given by    
\beqa
F_{\mu \nu}^a &=& \partial_\mu A_\nu^a - \partial_\nu A_\mu^a -g f^{abc} A_\mu^b
A_\nu^c \;,
\eeqa
with the coupling constant $g$ and the structure constants $f^{abc}$ of the gauge group $SU(N_c)$,
where $N_c$ is the number of colours. 
The covariant derivative in the fundamental representation of the gauge group is given by
\beq
D_\mu = \partial_\mu + igA_\mu \;, \label{covder}
\eeq 
with $A_\mu = A_\mu^a t^a$ and the $t^a$ are the generators of the gauge group.

Together the quark and gluon content of the action
\beq
S_{QCD} = \int d^4x \left(\bar{\Psi} \left( -\Dslash + m\right) \Psi + 
\frac{1}{4} F_{\mu \nu}^a F_{\mu \nu}^a\right)
\eeq 
ensures that QCD is invariant under local gauge transformations. Hereby it is assumed that
the path integral measure ${\cal D} [A \bar{\Psi} \Psi]$ is invariant by itself. 

The path integral in (\ref{genfunc0}) runs over all possible gauge field and Dirac
field configurations. This implies multiple counting of physically equivalent configurations,
{\it i.e.} configurations that are connected by a gauge transformation. Therefore the 
integration generates an infinite constant, the volume of the gauge group ${\cal G}$, which 
has to be absorbed in the normalisation. More important, the gauge freedom implies that the 
quadratic part of the gauge field Lagrangian has zero eigenvalues and therefore cannot be 
inverted. This prevents the definition of a perturbative gauge field propagator.

To proceed we introduce the notion of a gauge orbit $[A^g]$, which is a set of gauge configurations
that is related by a gauge transformation:
\beq
[A^g] := \left\{A^g = gAg^\dagger + gdg^\dagger : g(x) \in SU(N_c) \right\}. \label{orbit}
\eeq
All elements of a particular gauge orbit are physically equivalent. In order to single out 
one representative configuration from each orbit one has to impose a gauge fixing condition 
on the generating functional. This is conveniently done by the Faddeev-Popov 
procedure \cite{Faddeev:1967fc} (see also \cite{Pokorski:1987ed,Williams:2002dw} for 
pedagogical treatments of the subject). The idea is to insert the identity
\beq
1=\Delta[A^g] \int {\cal D}g \: \delta[F(A)] \label{FP}
\eeq
into the generating functional (\ref{genfunc0}). The gauge fixing condition $F(A)=0$ is 
supposed to be satisfied for one and only one gauge field configuration per gauge orbit. 
The 'Faddeev-Popov determinant' $\Delta[A^g]$ accounts for the functional determinant 
arising from the argument in the delta function. Problems with this gauge fixing 
procedure are discussed in more detail below. 

In linear covariant gauges the Faddeev-Popov determinant $\Delta[A]$ reads explicitly
\beq
\Delta[A^g] = \mbox{Det}\left(-\partial_\mu D_\mu^{ab}\right) ,
\eeq
where $D_\mu^{ab} = \partial_\mu \delta^{ab} + g f^{abc}A_\mu^c$ denotes the covariant derivative
in the adjoint representation. This determinant can be written as a functional integral 
over two new Grassmann valued fields $c$ and $\bar{c}$,
the so called 'Faddeev-Popov ghosts'. Also the gauge fixing condition $\delta[F(A)]$ can be 
represented by a Gaussian integral. We therefore finally arrive at the gauge fixed generating 
functional of QCD,
\beqa
\hspace*{-2cm}
Z[J,\sigma,\bar{\sigma},\eta,\bar{\eta}] &=& {\cal N}
\int {\cal D} [A \bar{\Psi} \Psi c \bar{c}]  
\exp\left\{\phantom{\int}\hspace*{-2mm} - S_{QCD}[A,\Psi,\bar{\Psi}] - S_{gf}[A,c,\bar{c}] \right. \nonumber\\
 && \hspace*{3cm} \left. + \int d^4x 
\left(A^a_\mu J^a_\mu+\bar{\eta}\Psi+\bar{\Psi}\eta + \bar{\sigma}c + \bar{c}\sigma 
\right) \right\}, 
\label{genfunc} 
\eeqa
with the gauge fixing part 
\beq
S_{gf} =  \int d^4x \left(\frac{\left(\partial_\mu A_\mu \right)^2}{2 \zeta} 
- i \partial_\mu \bar{c} D_\mu c\right) 
\label{Lagrangian}
\eeq
of the action. The first term in $S_{gf}$ stems from the gauge fixing condition and introduces
a gauge parameter $\zeta$, whereas the second
term represents the Faddeev-Popov determinant. The integral over the gauge group has been absorbed in
the overall normalisation ${\cal N}$. Note also that new sources $\sigma$ and 
$\bar{\sigma}$ for the antighost and ghost fields have been introduced.  
In the discussion below I frequently refer to Landau gauge, which is defined 
by the gauge condition $F(A)=\partial_\mu A_\mu=0$ and gauge parameter $\zeta=0$. The corresponding
gauge fixing condition of Coulomb gauge is $F(A)=\partial_i A_i=0$, where $i=1..3$ counts the spatial
components of the gauge field.

Although the action $S_{QCD} + S_{gf}$ is fixed wrt. local gauge transformations, there are two 
gauge symmetries left: The global gauge symmetry and the so called BRST-symmetry, which has been
found by Becchi, Rouet, Stora and Tyutin \cite{Becchi:1976nq,Iofa:1976je}. Since both will play 
an important role below, it is appropriate to discuss some of their properties.

The global gauge transformations of the gauge field and the quark field are given by 
\beqa
A_\mu &\rightarrow& A_\mu^\prime = e^{i t^a \Lambda^a} A_\mu e^{-i t^a \Lambda^a} \;, \\
\Psi &\rightarrow& \Psi^\prime = e^{i t^a \Lambda^a}\Psi \;,
\eeqa
with space-time independent parameters $\Lambda^a$ and the generators $t^a$ of the gauge group. Although
the global gauge transformation is a symmetry of the Lagrangian it is not immediately clear whether
a corresponding well defined charge exists, {\it i.e.} whether global gauge symmetry
is spontaneously broken or not. This issue will be discussed further below.

A BRST-transformation is formally equivalent to a local gauge transformation with the transformation
parameters $\Lambda^a(x)$ replaced by the product of a Grassmann number $\lambda$ and the ghost field, 
$\Lambda^a(x) \rightarrow \lambda c^a(x)$. Thus the transformation describes a global symmetry, 
since one is not free to treat different space-time points independently.
The infinitesimal transformations of the gluon, ghost 
and quark fields are given by\footnote{For brevity I only discuss the so called on-shell
transformations.} 
\beqa
s \Psi &=& -ig t^a c^a \Psi \;, \label{quark-brs}\\
s A_\mu^a &=&  D_\mu^{ab} c^b \;, \\
s c^a &=&  -\frac{g}{2} f^{abc} c^b c^c \;, \\
s \bar{c}^a &=&  0,
\eeqa
where $s$ denotes the Grassmann valued BRST-operator. This operator is nilpotent, 
{\it i.e.} $s^2=0$. Similar to global gauge symmetry, it is not clear whether the 
BRST-symmetry generates an unbroken BRST-charge $Q_B$ on the nonperturbative 
level of the path integral \cite{Fujikawa:1983ss,Baulieu:1998rp}. If so, then the
BRST-symmetry of the quantised, gauge fixed theory can be viewed as the analogue
to the gauge symmetry of the corresponding classical theory \cite{Kugo}.

We are now prepared for a discussion of three interesting ideas related to 
confinement: the Kugo-Ojima scenario, issues of positivity violations and the Gribov-Zwanziger
scenario.

 {\bf Kugo-Ojima scenario:}
 Confinement denotes the evidence that coloured particles have not been detected directly 
 in an experiment. On the theoretical side this corresponds to the absence of states with 
 non-zero colour charge from the physical part of the asymptotic state space of the theory. 
 A possible mechanism for this property of QCD has been proposed by Kugo and Ojima
 \cite{Kugo:1979gm}. Their scenario is explained in great detail in 
 \cite{Nakanishi:1990qm}, summaries can be found {\it e.g.} 
 in \cite{Alkofer:2000wg,Fischer:2003zc}. Here I give a brief account omitting 
 details. 
 
 The scenario may be phrased in terms of three lines of arguments:
 \begin{list}{A}{\setlength{\leftmargin}{9mm}}
 \item[(1.1)] If BRST-symmetry is an unbroken symmetry of gauge fixed, nonperturbative QCD,
              it can be used to define the physical part $\cal{W}_{\rm phys}$ of 
	      the state space $\cal{W}$ of QCD. 
 \item[(1.2)] If the global colour charge is unbroken, it can be shown that
              $\cal{W}_{\rm phys}$ contains colourless states only.
 \item[(1.3)] The cluster decomposition principle has to be violated in $\cal{W}$. 
 \end{list}
 These statements demand some explanations: \\
 Ad (1.1): in covariant gauges one encounters a 
 state space $\cal{W}$ that is equipped with an indefinite metric. This state space contains
 unphysical states ({\it e.g.} pure ghost or gluon states) as well as physical ones
 (bound states of quarks and/or gluons). An important step in a proof of confinement
 is to separate these. To this end Kugo and Ojima suggested to employ the nilpotent 
 BRST charge-operator $Q_B$, which is well-defined provided BRST-symmetry is unbroken. 
 The physical part $\cal{W}_{\rm phys} \subset \cal{W}$ of the state space of QCD is then defined 
 to contain exactly those states, which are nontrivially annihilated by $Q_B$.  
 It can be shown that $\cal{W}_{\rm phys}$ has a positive definite metric thus 
 allowing for a probabilistic interpretation of its expectation values. \\
 Ad (1.2): the second line of argumentation concerns the global colour charge.
 If this charge is well-defined ({\it i.e.} unbroken), one can show that 
 $\cal{W}_{\rm phys}$ contains colourless states only. This suggests that the corresponding 
 asymptotic state space to $\cal{W}_{\rm phys}$ represents the observable colour singlet 
 particles. Furthermore it can be shown, that the asymptotic states to the complementary 
 space of $\cal{W}_{\rm phys}$ do not contribute to physical S-matrix elements. 
 This so called 'BRST-quartet mechanism' ensures the absence of forward and backward 
 polarised gluons as well as ghost and antighost states from the physical spectrum of 
 the theory\footnote{The corresponding construction in QED is the familiar 
 Gupta-Bleuler formalism.}. A pedagogical treatment of this mechanism can 
 be found in \cite{Peskin:1995ev}. \\
 Ad (1.3): the cluster decomposition principle has to be violated in the total state space $\cal{W}$,
 but satisfied in $\cal{W}_{\rm phys}$. This ensures that all states in $\cal{W}_{\rm phys}$ 
 are 'localized' in the sense that one cannot detect possible unphysical components of 
 such a state. Within a meson state, for example, neither the quark nor the antiquark
 can then be detected individually. Nakanishi and Ojima argued that (1.3) is indeed satisfied 
 within the covariant operator formulation of QCD, see section 4.3.4 of \cite{Nakanishi:1990qm}.
 
 On the level of Green's functions there are some interesting relations between the
 statement (1.2) and the infrared behaviour of the propagators of Yang-Mills 
 theory. To understand the structure of this relation better we need to have a closer look
 on the global colour charge. The corresponding Noether current $J_\mu^a$ is given by
 \beq
  J_\mu^a = \partial^\nu F_{\mu \nu}^a + \left\{Q_B, D_\mu^{ab} \bar{c}^b \right\}, \label{charge}
 \eeq
 where $F_{\mu \nu}^a$ is the field strength tensor and $Q_B$ the BRST-charge operator.
 $D_\mu^{ab} \bar{c}^b$ denotes the covariant derivative of the anti-ghost field.
 From each of the two terms on the right hand side of (\ref{charge}) one can formally define a charge.
 In the following $G^a$ denotes the charge stemming from the first term whereas $N^a$ denotes the one from 
 the second term. These charges sum up to the global colour charge $Q^a$:
 \beq
  Q^a = G^a + N^a= \int d^3x \, \partial^\nu F_{0 \nu}^a + \int d^3x \,\left\{Q_B, D_0^{ab} \bar{c}^b \right\} .
 \eeq
 Depending on which of the three charges $Q^a$, $G^a$ and $N^a$
 are well-defined or spontaneously broken, Kugo, Ojima and Nakanishi suggested to
 distinguish three cases \cite{Kugo:1979gm,Nakanishi:1990qm}:
 \begin{itemize}
 \item[(i)] QED: In the Abelian theory the ghosts decouple and $N$ is always broken. 
  However, $G$ is broken as well due to the presence of massless photons in the field strength 
 tensor $F_{\mu \nu}$. One can then construct a linear combination of $N$ and $G$ such that 
 the massless contributions to both charges cancel. This combination defines an unbroken global 
 colour charge $Q$.
 \item[(ii)] Higgs-phase: In the Higgs-phase of the non-Abelian theory one has
 a vanishing charge $G^a$ for each massive gauge boson (since $\partial^\nu F_{0 \nu}^a$ is a 
 total derivative and the integral converges in this case). In these
 cases $N^a$ is spontaneously broken and therefore $Q^a$ is broken as well.
 \item[(iii)] Confinement phase: In the confinement phase of non-Abelian gauge theory
 $G^a$ and $N^a$ should be both unbroken and therefore combine to an unbroken global
 colour charge $Q^a$.
 \end{itemize}
 The last case is of interest to us. Massless asymptotic contributions to $G^a$ are 
 absent if the nonperturbative transverse gluon propagator is less singular than 
 a simple pole \cite{Nakanishi:1990qm}. The charge $G^a$ is then well-defined and vanishes
 (as a total derivative). Furthermore Kugo showed \cite{Kugo:1995km} that the charge $N^a$ is well-defined if the 
 dressed ghost propagator in Landau gauge is more singular in the infrared than a simple 
 pole. Thus these two conditions on the ghost and gluon propagators ensure an unbroken
 global colour charge $Q^a$ in QCD. This implies that $\cal{W}_{\rm phys}$ 
 contains colourless states only\footnote{The idea of the proof is easy to see: $\cal{W}_{\rm phys}$
 contains only states that are annihilated by $Q_B$. Furthermore $Q^a=N^a$ is proportional to $Q_B$. 
 Therefore the expectation value $\langle {\rm phys}| Q^a |{\rm phys} \rangle$ with 
 $|{\rm phys} \rangle \in \cal{W}_{\rm phys}$ vanishes, provided $Q_a$ is well defined.}. Evaluating the 
 infrared properties of the ghost and gluon propagator in Landau gauge thus enables 
 us to test (parts of) the Kugo--Ojima scenario. We will come back to this issue in 
 subsections \ref{IR} and \ref{prop}.
 
 {\bf Positivity violations:}
 The Kugo--Ojima scenario is one particular mechanism that ensures the
 probabilistic interpretation of the physical states of the quantum theory.  However, even if
 it were eventually shown not to be appropriate, it is apparent that
 there has to be {\em some} mechanism which singles out a physical, positive
 definite subspace in the state space of covariant QCD.  This suggests another criterion for
 confinement, namely violation of positivity.  If a certain
 degree of freedom has negative norm contributions in its propagator,
 it cannot describe a physical asymptotic state, {\it i.e.\/} there is
 no K\"all\'en--Lehmann spectral representation for its propagator.
 The precise mathematical structure of this condition in the context of 
 an Euclidean field theory has been formulated by Osterwalder and Schrader
 in the so called axiom of {\it reflection positivity} 
 \cite{Osterwalder:1973dx,Osterwalder:1975tc}. On the level of propagators
 this condition can be phrased as 
 \beqa 
  \Delta(t) &:=& \int d^3x \int \frac{d^4p}{(2\pi)^4}
  e^{i(t p_4+\vec{x}\cdot\vec{p})} \sigma(p^2) \\ 
  &=& \frac{1}{\pi}\int_0^\infty
  dp_4 \cos(t p_4) \sigma(p^2_4) \;\ge 0 \,,
 \label{schwinger} 
 \eeqa
 where $\sigma(p^2)$ is a scalar propagator function extracted from the respective propagator.
 We will see the consequences of this condition for the gluon and quark propagators in 
 more detail in subsections \ref{analytic} and \ref{analytic_quark}.

 {\bf Gribov-Zwanziger scenario:}
 A third interesting aspect of confinement is a scenario which is related to the 
 gauge fixing procedure. Since problems with gauge fixing are relevant for correlation functions,
 I discuss these briefly first and state the relation with confinement afterwards. To be
 specific I choose Landau gauge, since this is the gauge of choice for most of the results 
 discussed in this review. The situation is, however, similar in other gauges as for example
 Coulomb gauge.
 
\begin{figure}[t]
\begin{center}
\epsfig{file=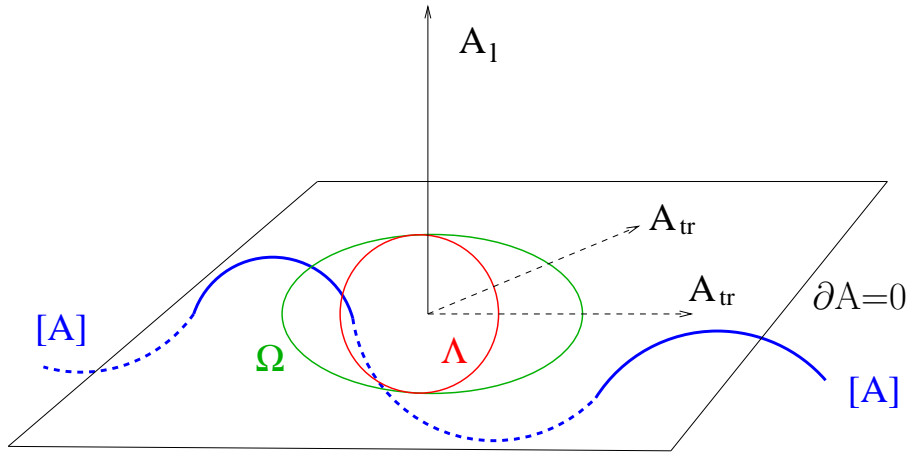,width=12cm}
\end{center}
\caption{Sketch of the hyperplane $\Gamma$ in gauge field configuration space 
obtained by Faddeev-Popov gauge fixing to Landau gauge, $\partial A = 0$. The transverse directions
of the gauge field, $A_{tr}$, generate $\Gamma$ whereas the longitudinal directions, $A_l$, are 
perpendicular to the hyperplane. Shown are furthermore the first Gribov region $\Omega$,
and the fundamental modular region $\Lambda$ containing the trivial configuration $A=0$. A gauge orbit
[A] intersects the hyperplane $\Gamma$ several times thus generating Gribov copies. The fundamental 
modular region $\Lambda$, however, is intersected only once.}
\label{FMR}
\end{figure}

 Fixing a gauge completely means singling out one representative configuration from each gauge 
 orbit (\ref{orbit}). 
 It has been shown by Gribov \cite{Gribov:1977wm} that the simple Faddeev-Popov 
 procedure is not sufficient in this respect. The Landau gauge condition, $\partial_\mu A_\mu = 0$,
 generates a hyperplane $\Gamma$ in gauge field configuration space which still contains gauge field
 configurations connected by a gauge transformation. These multiple intersection points of a gauge 
 orbit with $\Gamma$ are called {\it Gribov copies}, see figure \ref{FMR} for an illustration 
 (with $\partial_{\mu} A_\mu$ abbreviated by $\partial A$). 
 Gribov suggested to get rid of the 
 copies by restricting the hyperplane $\Gamma$ to the so called {\it Gribov region} $\Omega$. 
 This is conveniently done by minimising the following $L^2$-norm of the vector potential along 
 the gauge orbit:
\beq
\hspace*{-2.3cm} F_A(g) := ||A^g||^2 = ||A||^2 - 2 i \int d^4 x \; \tr(\omega \partial A ) 
+ \int d^4x \; \tr( \omega FP(A) \omega) + O(\omega^3),
\eeq  
with the gauge transformation $g=\exp(i\omega(x))$ and the Faddeev-Popov operator
\beq
FP(A) = -\partial D(A)^{ab} = -\partial^2 \delta^{ab}- g f^{abc}\partial_\mu A_\mu^c.
\eeq
 Any {\it local} minimum of this norm implements strictly the Landau gauge condition $\partial A=0$, 
 and restricts the Faddeev-Popov operator to positive eigenvalues. The Gribov region $\Omega$ 
 defined by this prescription has some interesting properties, which are discussed in detail 
 in \cite{Zwanziger:2003cf}. Important for us are: (i) $\Omega$ contains the origin of the gauge 
 field configuration space and therefore all configurations relevant for perturbation theory and 
 (ii) the lowest eigenvalue of the Faddeev-Popov operator $FP(A)$ approaches zero at the boundary 
 $\partial \Omega$, the {\it (first) Gribov horizon}. In general, there are still Gribov 
 copies contained in $\Omega$, therefore one needs to restrict the gauge field configuration
 space even further to the region of global minima of the $F_A$, which is called {\it fundamental 
 modular region}.  
 While a restriction to the Gribov region $\Omega$ in practical calculations can be achieved with 
 some effort, it is extremely cumbersome to identify the configurations of the fundamental modular 
 region. In other words, whereas local minima of $F_A$ are easily found it is almost impossible to 
 identify the global minima. Recent lattice studies 
 \cite{Cucchieri:1997dx,Mandula:1999nj,Alexandrou:2000ja,Alexandrou:2001fh,Alexandrou:2002gs,Giusti:2001xf,Silva:2004bv,Furui:2004cx,Sternbeck:2005tk,Lokhov:2005ra,Bogolubsky:2005wf} 
 concentrate some effort on evaluating the size of the problem. For the correlation functions
 of lattice gauge theory it seems that effects due to Gribov copies are more of quantitative and less
 qualitative nature. This situation may be even better for the continuum field theory: From an approach 
 using stochastic quantisation Zwanziger argued that Gribov copies inside the Gribov region do not 
 affect the Green's functions of the theory \cite{Zwanziger:2003cf}. This is highly fortunate, since 
 a restriction of the generating functional (\ref{genfunc}) to the Gribov region can be implemented in both
 lattice Monte-Carlo formulations and the DSE approach (see below).

 The basic idea of the Gribov-Zwanziger confinement scenario is the statement that gauge field 
 configurations close to the Gribov horizon drive the infrared properties of Yang-Mills theories 
 and are therefore responsible for colour confinement\footnote{In fact the relevant configurations 
 should be found close to the common boundary $\partial \Omega \cap \partial \Lambda$ of the Gribov 
 and the fundamental modular region. This is necessary to ensure equality of expectation values on 
 configurations within both regions.}. The scenario is probably most directly realized in Coulomb 
 gauge. In this gauge a renormalisation group invariant potential, the colour Coulomb potential, 
 can be identified \cite{Gribov:1977wm,Zwanziger:1998ez,Cucchieri:2000hv}. This potential is 
 an upper bound for the gauge invariant potential from the Wilson loop. Thus there is no confinement 
 without a confining Coulomb potential \cite{Zwanziger:2002sh}. On the other hand, the Coulomb 
 potential is related to an expectation value of the inverse Faddeev-Popov operator. 
 The presence of the Gribov horizon therefore triggers a growing potential and it has been shown
 that the potential is indeed (almost) linearly rising \cite{Szczepaniak:2001rg,Greensite:2003xf,Zwanziger:2003de,Feuchter:2004mk,Nakamura:2005ux}.

 In Landau gauge no corresponding quantity to the Coulomb potential has been identified so 
 far\footnote{Although there has been some progress to relate renormalisation group invariants 
 of Landau and Coulomb gauge \cite{Fischer:2005qe}.}. First lattice results for a specific 
 definition of a static quark potential show a flattening out of the potential at large 
 distances, which seems difficult to interpret in the absence of dynamical quarks 
 \cite{Nakamura:2005ux}. This issue remains to be clarified further.
 There are, however, other infrared effects in Landau gauge which are 
 argued to be driven by the presence of the Gribov horizon:
 \begin{itemize}
 \item[(I)] the presence of the Gribov horizon implies a dressed ghost propagator that is more 
 singular in the infrared than a simple pole \cite{Zwanziger:1993qr}.
 \item[(II)] the dressed gluon propagator has to vanish in the infrared \cite{Zwanziger:1991gz}.
 \end{itemize}
 These statements are known as Zwanziger's horizon conditions.
 They can be viewed as boundary conditions for any solutions of the properly 
 gauge fixed continuum gauge theory. The condition on the ghost is identical to the one encountered
 in the Kugo--Ojima scenario, whereas the condition on the gluon is stronger than the one discussed 
 above. We will come back to these conditions in subsections \ref{IR} and \ref{analytic}. 
 
 At the end of this subsection it is appropriate to give a very brief account on the 
 derivation of Dyson-Schwinger equations from the generating functional of a quantum field theory. 
 To keep the equations simple I will use a very dense, symbolic notation. Readers interested 
 in more details are referred to the textbooks \cite{Itzykson:1980rh,Rivers:1987hi} or the 
 reviews \cite{Roberts:1994dr,Alkofer:2000wg}. 
 Dyson-Schwinger equations follow from the generating functional (\ref{genfunc}) and the fact that the integral of
a total derivative vanishes, {\it i.e.}
\beqa
\hspace*{-1.8cm}
0 &=& \int {\cal D} [A \bar{\Psi} \Psi c \bar{c}] \frac{\delta}{\delta \phi} 
\exp\left\{- S_{QCD} - S_{gf} + \int d^4x\left(A J+\bar{\eta}\Psi+\bar{\Psi}\eta+\bar{\sigma}c+\bar{c}\sigma \right)\right\}
 \nonumber\\
\hspace*{-1.8cm}
&=&
\left\langle -\frac{\delta (S_{QCD}+S_{gf})}{\delta \phi} 
+ j \right\rangle  \label{genDSE}
\eeqa
for any field $\phi \in \{A,\Psi,\bar{\Psi},c,\bar{c}\}$ and its corresponding source 
$j \in \{J,\eta,\bar{\eta},\bar{\sigma},\sigma\}$. 
Equation (\ref{genDSE}) is valid provided that (i) a representation of the functional integral 
(\ref{genfunc}) exists and (ii) the measure ${\cal D}[A \bar{\Psi} \Psi c \bar{c}]$ is translationally 
invariant. In the following we will assume that these conditions are satisfied. Furthermore it is 
important to note that the form of (\ref{genDSE}) does not change when the functional integral
is restricted to the Gribov horizon (as required from the discussion above). The reason is that the 
Faddeev-Popov operator vanishes on the first Gribov horizon. Thus possible boundary terms from 
restricting the functional integral to the Gribov region vanish \cite{Zwanziger:2003cf}.

Further functional derivatives of the expression (\ref{genDSE})
with respect to a suitable number of fields $\phi$ and subsequently 
setting all sources to zero lead to the Dyson-Schwinger equation for any desired 
full n-point function. A similar procedure applied to the generating functional 
$W=\ln(Z)$ or the effective action $\Gamma = W + \langle\phi\rangle j $ leads to the 
DSEs for connected Green's functions and the ones for one-particle irreducible 
Green's functions.  An alternative derivation of DSEs employing Heisenberg's
equation of motion and equal time commutation relations can be found in \cite{Rivers:1987hi}.
Conceptual issues and results from Dyson-Schwinger equations in
Minkowski-space have been reviewed in \cite{Sauli:2004bx}.

Equation (\ref{genfunc}) and its derivatives constitute an infinite tower of coupled
integral equations. In the next subsection we will analyse some of these equations in more detail.

\setcounter{footnote}{0}  
\subsection{Infrared exponents of 1PI-Greens functions \label{IR}}
 
The study of infrared exponents of the propagators of Yang-Mills theory has been 
pioneered by Smekal, Hauck and Alkofer \cite{vonSmekal:1997is,vonSmekal:1998is} in
the late nineties. They were the first to realize an intricate interplay between the
ghost and gluon degrees of freedom, which led to the insight
that ghosts are dominant in the infrared. These results have been refined since
in a number of investigations 
\cite{Atkinson:1998zc,Atkinson:1998tu,Bloch:2001wz,Watson:2001yv,Zwanziger:2001kw,Zwanziger:2002ia,Zwanziger:2003cf,Lerche:2002ep,Schleifenbaum:2004id}.
Recently the infrared analysis of Yang-Mills theory has been extended to
include the whole tower of DSEs providing selfconsistent scaling laws for 
one-particle-irreducible (1PI) Green's functions with an arbitrary number of 
legs \cite{Alkofer:2004it}. In the following I summarize the general arguments in 
such an analysis (omitting technical details) and discuss the results. Consequences 
for the running coupling of Yang-Mills theory are discussed in the next subsection.

\begin{figure}[t]
\centerline{\epsfig{file=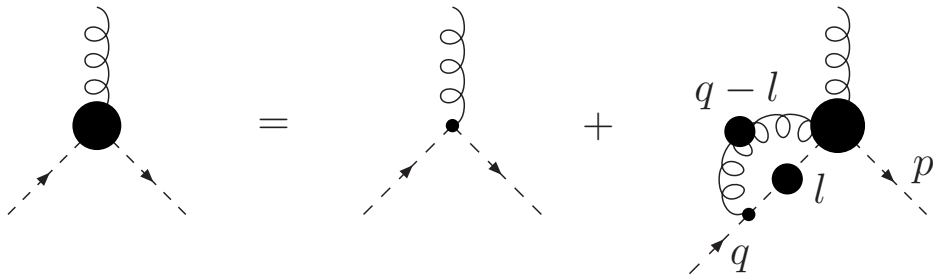,width=13cm}}
\caption{Dyson-Schwinger equation for the ghost-gluon vertex. The filled circles indicate dressed
Green's functions, {\it i.e.} propagators and vertices that contain all effects from quantum 
fluctuations. A wiggly line denotes a gluon propagator, whereas a dashed line stands for a ghost propagator.}
\label{DSE-ghg}
\end{figure}
The key starting point for an analysis of the infrared behaviour of 
1PI-Green's functions is the DSE for the ghost-gluon vertex. This equation is shown 
diagrammatically in figure \ref{DSE-ghg}. On the left hand side we have the 
dressed ghost-gluon vertex, which can be represented by  
\beq
\Gamma_\mu(q, p) = p_\mu E(p^2,q^2) + (p-q)_\mu F(p^2,q^2).
\eeq
Here $q,p$ are the momenta of the incoming and outgoing ghost and colour factors
have been omitted. The nonperturbative dressing functions $E(p^2,q^2)$ and $F(p^2,q^2)$
contain all effects from quantum fluctuations. The first diagram on the right hand side 
of the vertex DSE denotes the bare ghost-gluon vertex. It is given by
\beq
\Gamma^0_\mu(p, q) = \widetilde{Z}_1 p_\mu \,,
\eeq
where $\widetilde{Z}_1$ is the vertex renormalisation factor.
The second diagram on the right hand side contains a dressed four-point function
and a loop with dressed ghost and gluon propagators connected by a bare 
ghost-gluon vertex. This diagram has an interesting property: In Landau gauge, the 
momentum $q_\mu$ of the incoming ghost factorizes from the diagram. This can be
seen directly from figure \ref{DSE-ghg}: Since the gluon propagator $D_{\mu \nu}$ is 
transverse in Landau gauge, its contraction with the bare ghost-gluon vertex 
$\Gamma^0_\mu=l_\mu$ in the loop of the DSE gives 
$l_\mu D_{\mu \nu}(l-q) = q_\mu D_{\mu \nu}(l-q)$ and $q_\mu$ can be pulled out 
of the loop integral. Let us assume for the moment that\\
\hspace*{1cm}{\bf (A)} the loop-integral is finite in the infrared.\\ 
(We come back to this assumption in the paragraph below (\ref{IRsolution}).)
Since $q_\mu$ is factorized it is then clear that the dressing loop vanishes 
if $q_\mu$ goes to zero and the ghost-gluon vertex becomes bare in this limit.
The same can be shown if $p_\mu$ goes to zero. Thus the 
dressed ghost-gluon vertex in the infrared looks very much like a bare vertex.
This astonishing conclusion has been drawn by Taylor long ago \cite{Taylor:1971ff}
and has been confirmed recently by numerical studies of the ghost-gluon vertex on
the lattice and in the DSE-approach \cite{Cucchieri:2004sq,Sternbeck:2005qj,Schleifenbaum:2004id}.
Schleifenbaum {\it et al.} also investigated the mid-momentum behaviour of this vertex
and found only moderate structures in the vertex dressing \cite{Schleifenbaum:2004id}.

\begin{figure}[t]
\centerline{\epsfig{file=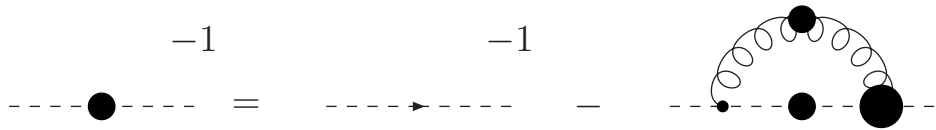,width=13cm}}
\caption{Dyson-Schwinger equation for the ghost propagator. On the left hand side is the inverse
of the dressed ghost propagator. On the right hand side we find the inverse of the
bare ghost propagator and a loop diagram containing dressed ghost and gluon propagators connected by 
one bare and one dressed ghost-gluon vertex. }
\label{DSE-gh}
\end{figure}

The simple structure of the ghost-gluon vertex has interesting consequences.
Consider the DSE for the ghost propagator, given in figure \ref{DSE-gh}. A bare or finite 
ghost-gluon vertex at small momenta admits a selfconsistent power law solution in the infrared: 
Writing the ghost and gluon propagators as
\beqa
D^G(p^2) &=& - \frac{G(p^2)}{p^2} \, , \nonumber\\
D_{\mu \nu}(p^2)  &=& \left(\delta_{\mu \nu} -\frac{p_\mu 
p_\nu}{p^2}\right) \frac{Z(p^2)}{p^2} \, ,
\eeqa
one finds power laws for the ghost and gluon dressing functions $G$ and $Z$ with
exponents given by\footnote{This can be checked easily by counting dimensions on both
sides of the equation. The loop-integral is dominated by momenta of the same magnitude 
as the external momentum. Thus, for small external momenta one can replace the 
propagators in the loop by their infrared approximation (\ref{kappa}). Both sides
of the equation are then proportional to $(p^2)^\kappa$.} \cite{vonSmekal:1997is,vonSmekal:1998is}
\beq
G(p^2) \sim (p^2)^{-\kappa}, \hspace*{1cm} Z(p^2) \sim (p^2)^{2\kappa}\,.
\label{kappa}
\eeq
The interesting point here is that selfconsistency forces an interrelation of the exponents 
such that they depend on one parameter $\kappa$ only. In this notation the Kugo-Ojima 
criterion (1.2) translates to the condition $\kappa > 0$, which means that the ghost 
propagator should be more singular and the gluon less singular than a simple pole. Zwanzigers 
horizon conditions state the same for the ghost propagator, but gives $\kappa > 0.5$ for 
the gluon dressing function. In a very general analysis of the ghost DSE Watson and Alkofer 
showed that the exponent $\kappa$ is positive \cite{Watson:2001yv}. The Kugo-Ojima criterion 
and the horizon condition (I) are therefore satisfied according to this analysis.

The specific value for $\kappa$ depends on the details of the dressing of the ghost-gluon vertex
at small momenta. For a range of possible dressings this has been investigated by Lerche and
Smekal in \cite{Lerche:2002ep}. They argued that $0.5 \le \kappa < 0.6$ with a possible upper limit
given by the result for a bare ghost-gluon vertex,  $\kappa = (93 - \sqrt{1201})/98 \approx 0.595$,
a value independently found also in \cite{Zwanziger:2001kw}. Recent investigations in the 
framework of exact renormalisation group equations confirm this 
range \cite{Pawlowski:2003hq,Fischer:2004uk}. Apart from the lower bound all possible values 
of $\kappa$ lead to a vanishing gluon propagator in the infrared in agreement with 
Zwanzigers horizon condition (II). 

\begin{figure}[t]
\centerline{\epsfig{file=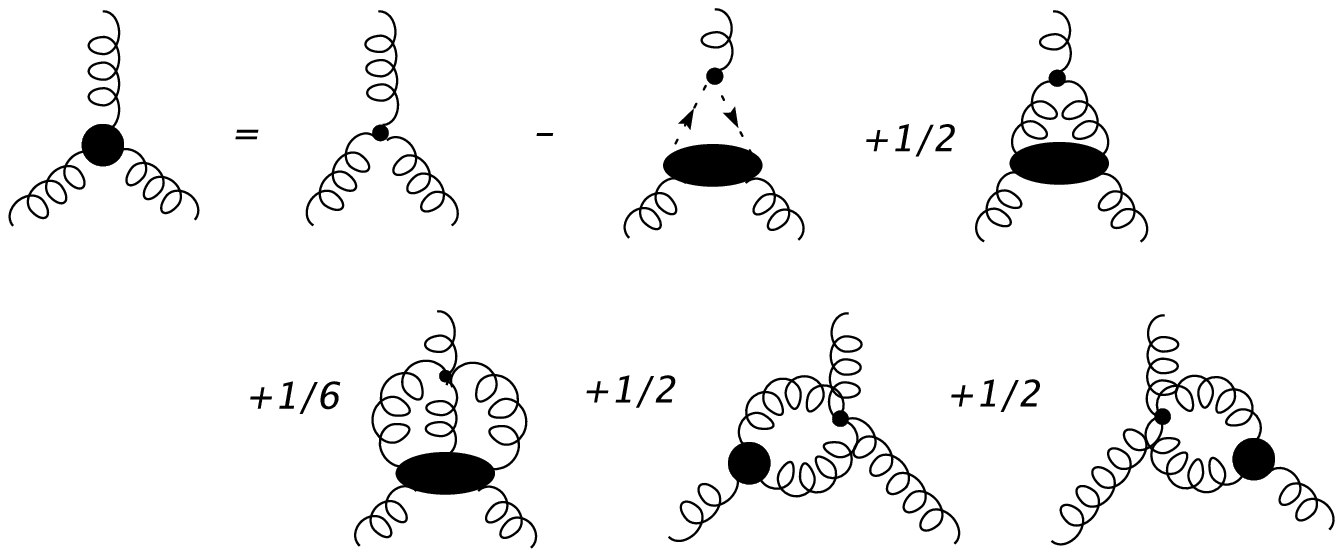,width=13cm}}
\vspace*{8mm}
\centerline{\epsfig{file=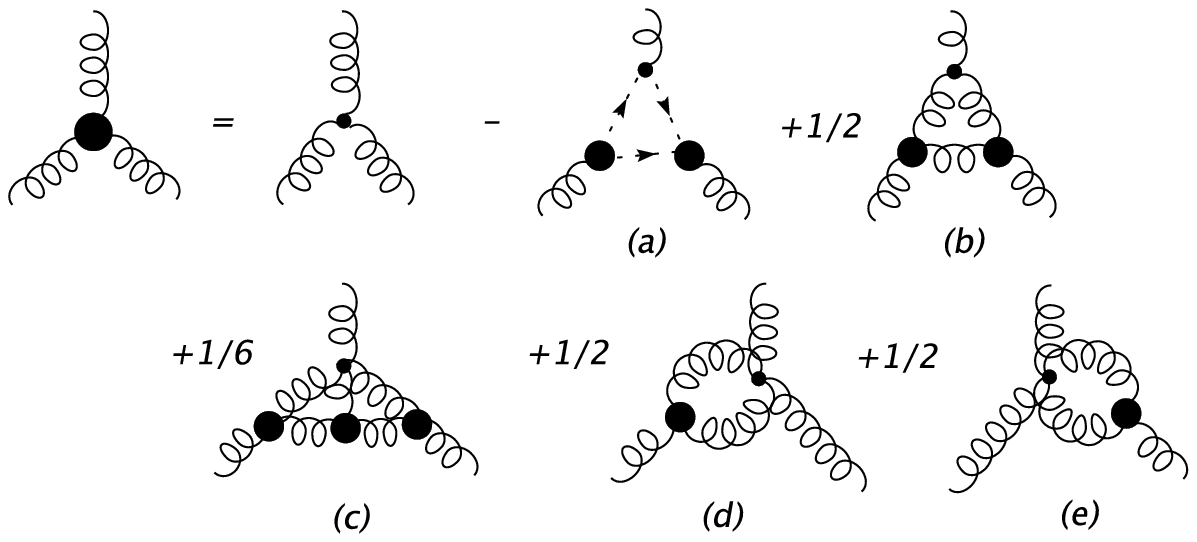,width=13cm}}
\caption{Exact Dyson-Schwinger equation for the three-gluon vertex (upper equation)
and lowest order in a skeleton expansion of the four-
and five-point functions (lower equation). All internal propagators in the diagrams are to
be understood as fully dressed.} 
\label{DSE-3g}
\end{figure}

On the basis of the relation (\ref{kappa}) one can also determine the
infrared exponents of higher n-point functions. The key idea here is 
to solve the corresponding DSEs 
order by order in a skeleton expansion ({\it i.e.} a loop expansion using 
dressed propagators and vertices). This program has been carried out by Alkofer, Fischer
and Llanes-Estrada in \cite{Alkofer:2004it}. It turns out that in this expansion the 
Green's functions can only be infrared singular, if all external scales 
go to zero. Thus to determine the degree of possible singularities it is 
sufficient to investigate the DSEs in the presence of only one external 
scale $p^2 \ll \Lambda^2_{QCD}$, where $\Lambda_{QCD}$ is of the order of a view hundred MeV.
As an example consider the DSE for the three-gluon 
vertex. In figure \ref{DSE-3g} we see the
full equation as well as an approximation in the lowest order of a skeleton 
expansion. In the presence of one (small) external 
scale the approximated DSE has a selfconsistent power law solution given 
by\footnote{Again this can be seen by counting anomalous dimensions on both
sides of the equations. The loops are dominated by momenta of the same magnitude as the 
external scale, thus one can substitute the propagators and vertices in the loops
by their infrared scaling laws. The leading diagram on the right hand side is the
one involving ghosts, diagram (a), the others are less singular (recall $\kappa > 0$). 
The diagram (a) is proportional to $(p^2)^{-3\kappa}$.}
\beq
\Gamma^{3g}(p^2) \sim (p^2)^{-3\kappa}. \label{IR_3g}
\eeq
The vertex is strongly singular in the infrared. One can see by induction that this 
solution is also present if terms to arbitrary high order in the skeleton expansion are taken 
into account. Thus the skeleton expansion is stable wrt. the infrared solution
of the DSEs. 

This technique can be applied to any other DSE as well. A 
self-consistent solution of the whole tower of DSEs is then 
given by \cite{Alkofer:2004it}
\beq
\Gamma^{n,m}(p^2) \sim (p^2)^{(n-m)\kappa}. \label{IRsolution}
\eeq
Here $\Gamma^{n,m}(p^2)$ denotes the infrared leading dressing function of 
the 1PI-Green's function with $2n$ external ghost legs and $m$ external 
gluon legs. By counting anomalous dimensions it can be checked that
the expression (\ref{IRsolution}) does not only solve the approximated
but also the full three-gluon vertex DSE in figure \ref{DSE-3g} selfconsistently. 
Furthermore, inserting 
$\Gamma^{1,2}(p^2) \sim (p^2)^{-\kappa}$ together with the power laws 
(\ref{kappa}) into the DSE for the ghost-gluon vertex, figure \ref{DSE-ghg},
one can verify the basic assumption {\bf (A)} from the beginning of this subsection:
the loop-integral of the vertex dressing is indeed finite in the infrared. 
Thus (\ref{IRsolution})
is a truly selfconsistent infrared solution of the tower of DSEs\footnote{It 
is worth mentioning that the solution (\ref{IRsolution}) also has the
correct scaling behaviour such that the Slavnov-Taylor identities of
the renormalisation constants are satisfied. Since the theory is 
multiplicative renormalizable these functions scale with the 
renormalisation point $\mu^2$ in the same way as the 1PI-functions with the
external scale $p^2$. E.g. the relation $Z_1/Z_3 = \widetilde{Z}_1/\widetilde{Z}_3$
between the three-gluon vertex, gluon propagator, ghost-gluon vertex and ghost 
propagator renormalisation constant leads to $Z_1(\mu^2) = (\mu^2)^{-3\kappa}$, 
which agrees with (\ref{IR_3g}).}.

An important aspect of the selfconsistent solution (\ref{IRsolution}) is the 
observation that diagrams containing ghost-loops dominate the infrared behaviour 
of every DSE. On the level of the generating functional (\ref{genfunc})
this corresponds to the statement that the Faddeev-Popov determinant dominates
the quantum fluctuations in the infrared and one can define an infrared asymptotic 
theory by neglecting the Yang-Mills action, {\it i.e.} setting 
$[\exp(-S_{YM})]=1$ \cite{Zwanziger:2003cf}. The solution of this theory is given
by the power laws (\ref{IRsolution}) (in the presence of only one external scale).
Interestingly, this limit is a continuum analogue of the strong coupling limit 
of lattice gauge theory. It still persists even if quarks are included, as will 
become clear in subsection \ref{unquench}. Zwanziger showed that there is an 
infinite mass gap in this asymptotic theory \cite{Zwanziger:2003cf}. On this basis 
he suggested a picture of confinement in Landau gauge: Although the
infrared modes of the gauge field are suppressed (vanishing gluon propagator), its
ultraviolet modes fluctuate wildly because $[\exp(-S_{YM})]=1$. This causes the 
decoherence of any field that carries colour; the corresponding particles do not 
propagate and are therefore confined. In full QCD in Landau gauge 
it is then the fluctuations of the gauge field around $\Lambda_{QCD}$ that should be 
responsible for confinement. 
Possible candidates for gauge field configurations 
with this property have been identified in the SU(2)-theory on the lattice: In 
\cite{Langfeld:2001cz} it was shown that much of the strength of the gluon propagator 
in this region vanishes if center vortex configurations were eliminated from the 
statistical ensemble. The reduced ensemble also does not reproduce the linear 
rising static quark-antiquark potential and is therefore not confining in contrast 
to the full ensemble including center vortices. A detailed discussion of the confining
role of center vortices can be found in \cite{Greensite:2003bk}.

Finally there is caveat: it it necessary to keep in mind that selfconsistency is not 
enough to establish (\ref{IRsolution}) as the 'true' solution of Yang-Mills 
theory in the infrared. There may be other selfconsistent solutions of the DSEs. In 
this case one needs other criteria to decide which solution is the one realized in 
nature. However, besides its relations to confinement there is a further interesting
property of the solution (\ref{IRsolution}) that qualifies it as a promising 
candidate: it leads to qualitative universality of the running coupling in the 
infrared. This is the subject of the next subsection.

\setcounter{footnote}{0}  
 \subsection{The infrared behaviour of the running coupling \label{coupling}}

The infrared behaviour of the running coupling of Yang-Mills theory has been 
investigated in a number of approaches. In the past years evidence is growing that 
the old picture of 'infrared slavery', {\it i.e.} the notion of an infrared singular 
behaviour of the running coupling is not appropriate. Instead, evidence from 
continuum field theory suggests that the running coupling freezes below 
a certain momentum scale and develops a fixed point at $p^2 \rightarrow 0$. This 
behaviour has been conjectured from phenomenological investigations (see {\it e.g.} 
\cite{Baldicchi:2002qm,Brodsky:2002nb,Grunberg:2006jx} and 
references therein). It is also found in 'optimized perturbation theory' 
\cite{Mattingly:1993ej} as well as in 'analytic perturbation theory' 
\cite{Shirkov:1997wi,Shirkov:2001sm,Howe:2003mp,Nesterenko:2004tg,Milton:2005hp}.
Certainly, since the infrared limit of QCD is concerned here, genuine nonperturbative
approaches seem to be mandatory to underpin these findings. It is therefore quite
satisfactory that a fixed point behaviour of the running coupling has also been found
in the framework of the exact renormalisation group 
\cite{Gies:2002af,Pawlowski:2003hq,Fischer:2004uk} and from solutions of Dyson-Schwinger
equations. In fact the first nonperturbative result stems from Smekal, Hauck and Alkofer
within the DSE-approach \cite{vonSmekal:1998is} and has since been elaborated further
by Lerche and Smekal \cite{Lerche:2002ep} and Alkofer, Fischer and Llanes-Estrada
\cite{Alkofer:2004it}. In the following I will first outline what we find from DSEs 
and then comment on relations to other approaches.

In Landau gauge renormalisation group invariant couplings can be defined from 
either of the primitively divergent vertices of Yang-Mills-theory, 
{\it i.e.} from the ghost-gluon vertex ($gh$), the three-gluon vertex ($3g$) or 
the four-gluon vertex ($4g$) via
\beqa
\alpha^{gh}(p^2) &=& \frac{g^2}{4 \pi} \, G^2(p^2) \, Z(p^2),                \label{gh-gl}  \\
\alpha^{3g}(p^2) &=& \frac{g^2}{4 \pi} \, [\Gamma^{0,3}(p^2)]^2 \, Z^3(p^2), \label{3g}     \\
\alpha^{4g}(p^2) &=& \frac{g^2}{4 \pi} \, [\Gamma^{0,4}(p^2)]^2 \, Z^4(p^2). \label{4g}
\eeqa
Here $\Gamma^{0,3}$ denotes the infrared leading dressing of the three-gluon-vertex 
and $\Gamma^{0,4}$ the corresponding one for the four-gluon vertex. Details on the 
derivation of these expressions are given in \cite{Alkofer:2004it}.
Note that the multiplicity of the various dressing functions correspond to the
multiplicity of the legs of each vertex. Since the ghost-gluon vertex is a finite object in Landau
gauge its coupling does not depend on the vertex dressing function. This coupling can 
therefore be determined from the propagators of Yang-Mills theory alone.

Before we discuss the infrared behaviour of these couplings let us recall some 
of their general properties. First, it is important to note that the definitions 
(\ref{gh-gl})-(\ref{4g}) correspond to a momentum subtraction scheme (at a symmetric
Euclidean momentum point). This entails that
couplings from different vertices do not necessarily agree with each other (unlike in 
the $\overline{MS}$-scheme), but are related via Ward-identities of the 
renormalisation functions \cite{Celmaster:1979km,Pascual:1980yu}. Although on the one-loop 
level these differences are rather small \cite{Celmaster:1979km}, to my knowledge 
there is no argument for this to persist to higher loop order or in the nonperturbative
framework discussed here. 
Second, perturbation theory suggests that the gauge dependence of these couplings is
rather weak in the vicinity of Landau gauge. This may be related to the fact that
Landau gauge is a fixed point under the renormalisation group flow \cite{Ellwanger:1995qf},
and justifies Landau gauge as a good starting point for the investigation of these 
couplings. Below I will comment on corresponding results in a more general class 
of transverse gauges.

Using the DSE-solution (\ref{IRsolution}) in place of the various dressing
functions it is easy to see that all three couplings approach a fixed 
point\footnote{In the literature the existence of such a fixed point has been attributed 
frequently to the dynamical generation of a gluon mass, see {\it e.g.} \cite{Nesterenko:2004tg}
and references therein. However, from (\ref{gh-gl2})-(\ref{alpha2}) we see that
this is by no means a necessity. In the language of the infrared exponent 
$\kappa$ a gluon mass would correspond to the special case of $\kappa=0.5$. 
The mere existence of the fixed points (\ref{gh-gl2})-(\ref{alpha2}) however, 
does {\it not} depend on any special value of $\kappa$. Only its value does, 
see the discussion around (\ref{fixed}).} in 
the infrared \cite{Alkofer:2004it}:  
\beqa
\alpha^{gh}(p^2) &=& \frac{g^2}{4 \pi} \, G^2(p^2) \, Z(p^2) 
     \hspace*{7mm} \stackrel{p^2 \rightarrow 0}{\sim} \hspace*{0mm} 
     \frac{c_1}{N_c} \,, \label{gh-gl2}\\
\alpha^{3g}(p^2) &=& \frac{g^2}{4 \pi} \, [\Gamma^{0,3}(p^2)]^2 \, Z^3(p^2) 
    \hspace*{0mm} \stackrel{p^2 \rightarrow 0}{\sim}
     \hspace*{0mm} \frac{c_2}{N_c} \,,\\
\alpha^{4g}(p^2) &=& \frac{g^2}{4 \pi} \, [\Gamma^{0,4}(p^2)]^2 \, Z^4(p^2) 
    \hspace*{0mm} 
    \stackrel{p^2 \rightarrow 0}{\sim} \hspace*{0mm} \frac{c_3}{N_c} \,.
     \label{alpha2}
\eeqa
They are thus qualitatively universal in the infrared.
As explained above, the constants $c_{i=1..3}$ may be different for each coupling and depend in 
particular on the respective choice of the tensor component used to extract the 
vertex dressing functions $\Gamma$. For the coupling (\ref{gh-gl2}) of the ghost-gluon vertex 
this fixed point can be explicitly calculated from the coupled set of DSEs for the ghost 
and gluon propagator. Employing a bare ghost-gluon vertex Lerche and Smekal found \cite{Lerche:2002ep}
\beq
\alpha^{gh}(0) = \frac{2 \pi}{3 \, N_c} \frac{\Gamma(3-2\kappa) \; \Gamma(3+\kappa) \; \Gamma(1+\kappa)}
{\Gamma^2(2-\kappa) \; \Gamma(2\kappa)} \approx 8.92/N_c. \label{fixed}
\eeq
with $\kappa \approx 0.596$. The dependence of the fixed point on the exponent $\kappa$ is rather weak. 
With  $0.5 \le \kappa <0.6$ one obtains roughly $2.5 < \alpha(0) <3$ for $N_c=3$ \cite{Lerche:2002ep}.
The expressions for the other two couplings involve vertex dressing functions which receive
contributions from all orders of the (nonperturbative) skeleton expansions of their DSEs 
(see last subsection). Unfortunately this makes it extremely difficult to determine their 
fixed point values and no quantitative statement can be made at present. 

This result, an infrared fixed point of the running coupling, is interesting for several
reasons. The integration over a bounded running coupling is finite, which simplifies the 
calculation of many observables. Moreover, as discussed {\it e.g.} in \cite{Brodsky:2003rs}, 
elements of conformal field theory then become relevant at small 
momentum transfers and commensurate scale relations between different observables hold. 
These relations link experimental observables to each other without any renormalisation scale 
or scheme ambiguities. They employ so called {\it effective charges} \cite{Grunberg:1982fw}, 
which are defined directly from observables. These charges are analytical and non-singular 
by definition. The couplings (\ref{gh-gl})-(\ref{4g}), however, are defined directly 
from the vertices of the theory. It is thus a highly nontrivial result that they are 
also non-singular and analytic along the positive Euclidean momentum axis. This certainly 
suggests some relation between these two types of couplings which would be interesting to 
clarify in the future. A direct comparison between an effective charge extracted 
from the Bjorken sum rule and the coupling from the ghost-gluon vertex has been performed 
in \cite{Deur:2005cf}. Both charges have an infrared fixed point and agree well even 
quantitatively. It is, however, not yet understood whether this agreement signals a 
relation between these couplings or whether it is purely accidental.
 
There is a further interesting aspect on commensurate scale relations. Since they are based
on gauge invariant effective charges they hold in any gauge. Thus, if a relation between
couplings from the vertices of the theory and effective charges exists, one should find
an infrared fixed point for the former type of couplings in other gauges as well. This 
question has been addressed recently by Fischer and Zwanziger in \cite{Fischer:2005qe}. 
We solved the coupled system of DSEs for the ghost and gluon propagator in the infrared 
in a class of transverse gauges that interpolate between Landau and Coulomb gauge.
From the ghost-gluon vertex in these gauges we found two invariant running couplings,
which both survive in the Coulomb gauge limit. One of these becomes the colour 
Coulomb potential. The other coupling indeed has a fixed point in the whole class of 
transverse gauges including Coulomb gauge. The value of this fixed point is the same as
in Landau gauge, $\alpha^{gh}(0) = 8.92/N_c$, for all transverse gauges but the Coulomb 
gauge limit. A very recent calculation of Schleifenbaum, Leder and Reinhardt 
finds a substantial increase of this value in the Coulomb gauge 
limit \cite{Schleifenbaum:2006bq}. It would  
certainly be interesting to perform a similar analysis for the couplings from the 
three-gluon and four-gluon vertices. One may hope that there are gauge invariant 
features (at least within certain classes of gauges) in all couplings defined from the 
vertices of Yang-Mills theory\footnote{In this respect it is interesting to note that a 
well-defined class of gauges exists that interpolates also from Landau gauge towards
maximal Abelian gauge \cite{Dudal:2005zr}, where confinement may be explained via a dual 
superconductor scenario. First results on the infrared behaviour of ghost and gluon 
propagators in these gauges are reported in \cite{Capri:2006vv}, while results for the
coupling are not yet available.}.
 
\subsection{Numerical solutions compared to results from lattice calculations \label{prop}}

In the last two subsections we discussed some of the infrared properties of one-particle
irreducible Green's functions and related definitions of the running coupling.
We are now ready to establish the connection between these results at very
small momenta and those known from perturbation theory at large momenta. To this
end we will discuss results from numerical solutions to the DSEs, which connect
the perturbative and nonperturbative regime. At intermediate momenta these
results can be compared to corresponding lattice calculations. We will
see that overall both approaches agree well. However, we will also find
some interesting differences. In subsection \ref{IR}, we saw that the gluon propagator
vanishes in the infrared. Most contemporary lattice calculations obtain a finite propagator 
at zero momentum. This is an open problem, which is discussed frequently in the literature 
\cite{Fischer:2005ui,Silva:2005hb,Sternbeck:2005tk,Boucaud:2005ce,Cucchieri:2006za}. At the 
end of this subsection I will outline a possible route towards an explanation of this difference in 
terms of boundary conditions and finite volume effects on the compact lattice manifold.

\begin{figure}[t]
\centerline{\epsfig{file=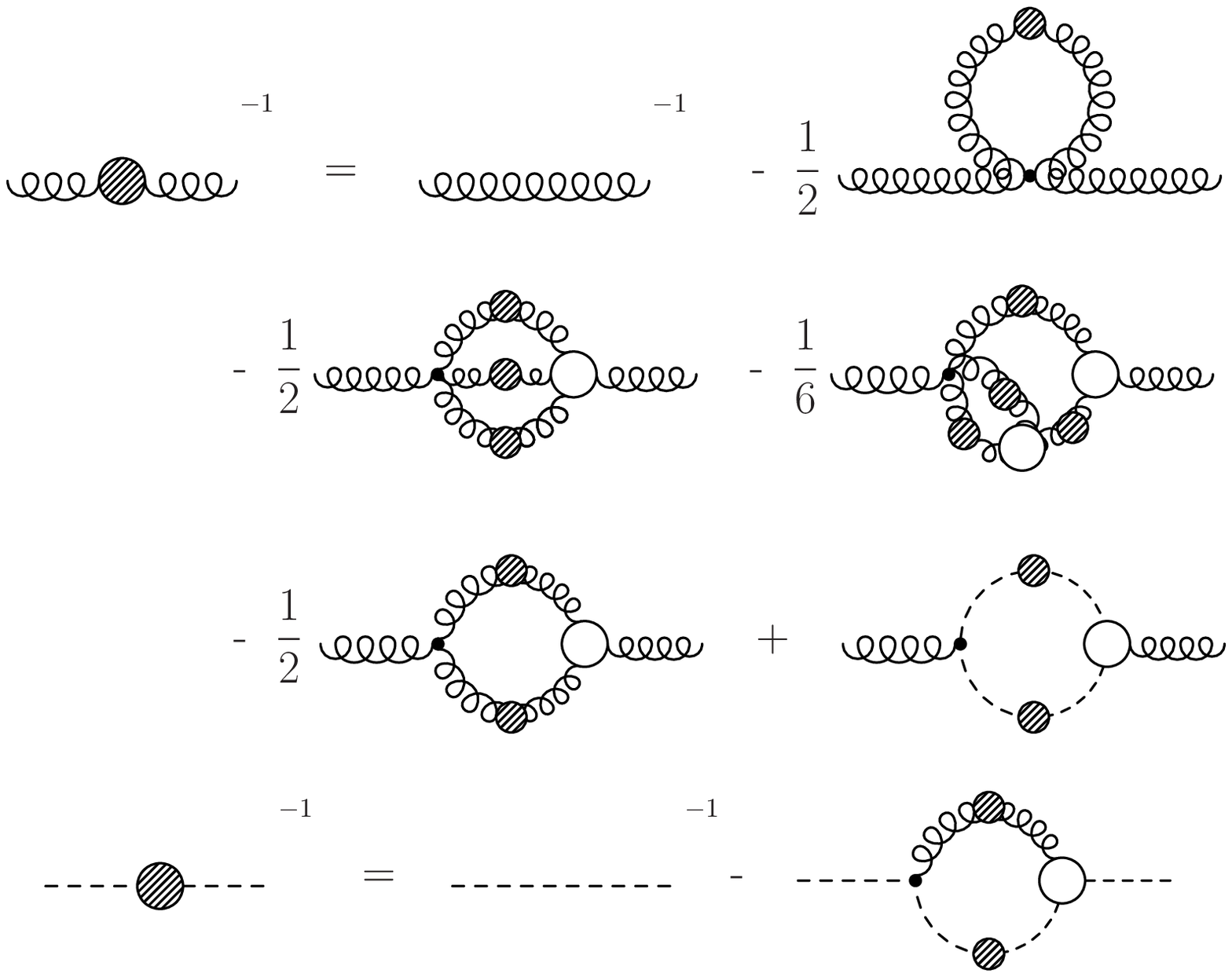,width=12cm}}
\caption{Dyson-Schwinger equations for the gluon and ghost propagator. Filled circles denote dressed 
propagators and empty circles denote dressed vertex functions.}
\label{DSE-gl}
\end{figure}

The DSEs for the ghost and gluon propagator are given diagrammatically in 
figure \ref{DSE-gl}. They form a coupled system of equations which demand  
dressed ghost-gluon, three-gluon and four-gluon vertices as input. Assuming ansaetze for
the vertices these equations have been solved selfconsistently for the first time
by Smekal, Hauck and Alkofer in \cite{vonSmekal:1997is,vonSmekal:1998is}. 
Since then numerical techniques have been improved \cite{Atkinson:1998tu} and 
the technique of angular approximation of loop integrals has been replaced by 
full fledged numerical integration methods \cite{Fischer:2002hn,Bloch:2003yu,Fischer:2005en}. 
These contemporary solutions still have the same qualitative structure as the 
ones of \cite{vonSmekal:1997is,vonSmekal:1998is}, but are greatly improved
in terms of quantitative reliability. 

It is evident that the quality of the solutions of the DSEs depends on the
quality of the vertex truncation. The situation is best for the ghost-gluon vertex.
Evidence from the analytic infrared analysis ({\it cf.} subsection \ref{IR}), from 
lattice calculations (although yet only available for specific kinematics) 
\cite{Cucchieri:2004sq,Sternbeck:2005qj} and a semi-selfconsistent numerical solution 
of the vertex DSE \cite{Schleifenbaum:2004id} suggest that a bare vertex approximation 
is justified. Mild effects from neglecting nontrivial vertex dressing should show up only
in the mid-momentum region. This is also the region where the selfinteraction of the
gluon is least known. Lattice studies of the three-gluon vertex 
\cite{Alles:1996ka,Boucaud:2002fx} cover only specific kinematical regions and cannot 
yet be used to constrain an ansatz for this vertex\footnote{There is some evidence
from the lattice that the vertex is diverging in the infrared \cite{Boucaud:2002fx}
as predicted from our infrared analysis in subsection \ref{IR}. However the data do not 
allow a determination of the precise strength of this divergence.}. 
In practise ansaetze for this vertex have been 
used which agree with the infrared analysis in \ref{IR}. They also lead to the correct behaviour 
of the propagators in the ultraviolet according to resummed perturbation theory. 
Contributions involving the four-gluon vertex have been neglected so far\footnote{A first 
attempt to include the gluon two-loop diagrams has been made in \cite{Bloch:2003yu}.
However, the employed ansaetze for the three- and four-gluon vertices disagree 
with the requirement of selfconsistency in the vertex DSEs. Unfortunately this 
sheds some doubt on the conclusiveness of the results achieved there.}.
While at first sight this last choice might seem arbitrary, it is in fact well justified
from the infrared analysis in subsection \ref{IR}: the gluon two-loop diagrams are 
subleading in the infrared. They are 
also subleading in the ultraviolet, since they do not appear to leading order in
perturbation theory. In the intermediate momentum regime such a truncation together 
with uncertainties in the dressing of the three-gluon and ghost-gluon vertex introduces 
quantitative errors. These have to be controlled a posteriori by a comparison of the results 
for the gluon and ghost propagator with corresponding lattice calculations. 

\begin{figure}[t]
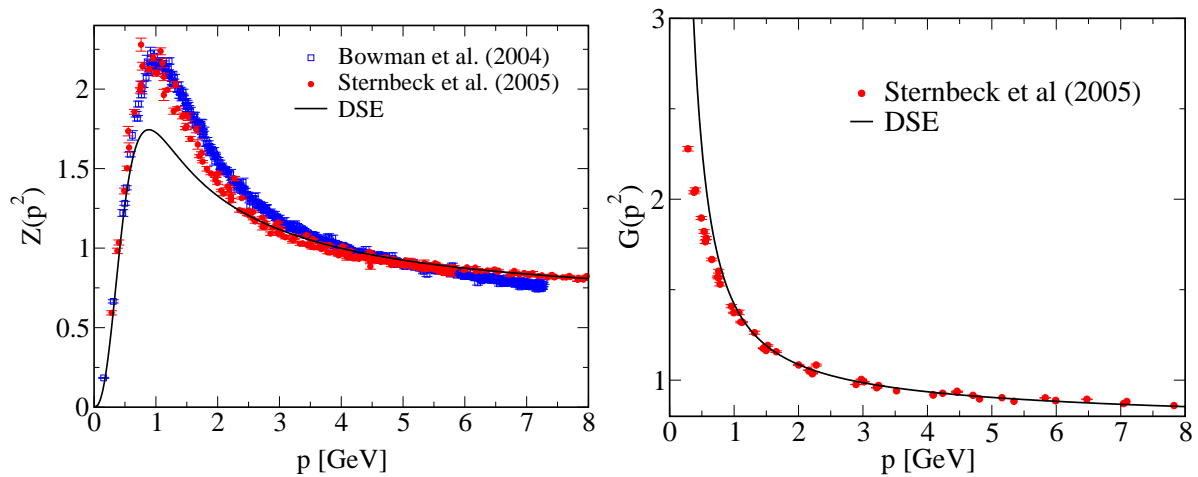

\centerline{
\epsfig{file=figure6a.eps,width=7.8cm}
\hfill
\epsfig{file=figure6b.eps,width=7.8cm}}
\caption{Results for the gluon and ghost propagator from Dyson-Schwinger equations 
in the continuum compared with recent lattice data. 
From: Fischer and Alkofer \cite{Fischer:2002hn} (DSE),
Bowman {\it et al.} \cite{Bowman:2004jm} (lattice),
Sternbeck {\it et al.} \cite{Sternbeck:2005tk}) (lattice).    
\label{ghostglue}}
\end{figure}
\begin{figure}[t]
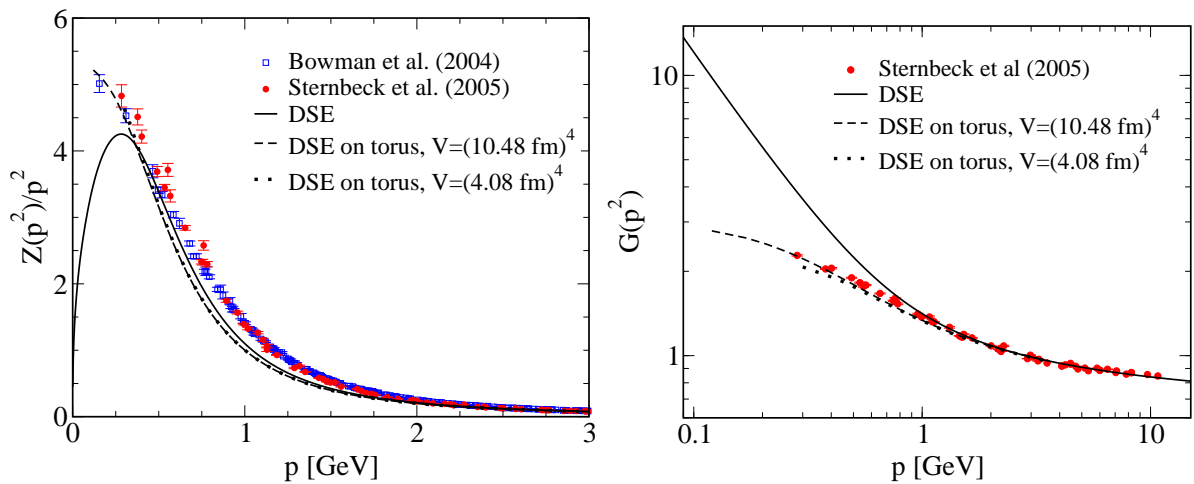

\centerline{
  \epsfig{file=figure7a.eps,width=7.8cm}
  \hfill
  \epsfig{file=figure7b.eps,width=7.8cm}}
\caption{Same results as in figure \ref{ghostglue} but displayed differently. In 
addition results from DSEs on a torus are shown. 
From: Fischer and Pennington, \cite{Fischer:2005nf} (DSE on torus).
\label{ghostglue-vol}}
\end{figure}

Such a comparison can be seen in figure \ref{ghostglue}. The global structure of the solutions in both
the lattice and the DSE approach is the same. In the ultraviolet all solutions reproduce
perturbation theory as expected (this can be shown analytically for the DSEs). 
In the intermediate momentum region one observes a larger bump in the gluon dressing 
function from the lattice compared to the DSE result. This difference serves as an 
estimate of the neglected effects due to the gluon selfinteraction. In the infrared 
both approaches seem to agree nicely 
on a linear plot. The numerical results from the DSEs agree here with
the analytical results discussed in subsection \ref{IR} (this can be seen best
on a log-log plot, displayed {\it e.g.} in \cite{Fischer:2002hn}) and the lattice
data are not far away. However, we have to look closer here.

In figure \ref{ghostglue-vol} we see the same results displayed differently. Plotted is 
the gluon propagator $Z(p^2)/p^2$ instead of the dressing function $Z(p^2)$.
The ghost dressing function $G(p^2)$ is shown on a log-log plot. 
In the infrared we
clearly see differences now. The ghost dressing function in figure \ref{ghostglue-vol} diverges
in the infinite volume/continuum limit, whereas it stays finite on the lattice,
{\it i.e.} on the compact manifold. For the gluon propagator, the differences can be 
expressed in terms of the infrared power law, 
\beq
Z(p^2) \sim (p^2)^{2\kappa}.\label{Z_IR} 
\eeq
One obtains
$\kappa \approx 0.5$ (IR-finite) on a compact manifold, whereas $\kappa \approx 0.6$ 
(IR-vanishing) on $R^4$ in agreement with the analytical results ({\it cf.} subsection 
\ref{IR})\footnote{The only recent numerical DSE-study that finds an infrared finite 
gluon propagator also in the continuum is reported in \cite{Aguilar:2004sw}. This 
study, however, is hardly conclusive, since it employs a (highly dubious) 
renormalisation prescription that enforces the propagator to be constant in the infrared.}.
This is a decisive difference, since it can be shown that an infrared vanishing 
gluon propagator cannot have a positive definite spectral function and is therefore
confined, see next subsection. Indeed, Zwanziger has argued that the lattice 
gluon propagator should vanish in the continuum limit \cite{Zwanziger:1991gz}, 
{\it cf.} subsection \ref{conf}. However, no statement could be made as to the rate 
with which the continuum limit behaviour is approached. Current extrapolations of 
lattice data to the infinite volume limit are under discussion
\cite{Bonnet:2001uh,Silva:2005hb,Tok:2005ef,Cucchieri:2006za,Boucaud:2006pc}.
 
As an attempt to clarify this situation we changed the base manifold on our DSEs from $R^4$ to
the lattice manifold, {\it i.e.} a torus with periodic boundary 
conditions \cite{Fischer:2002eq,Fischer:2005ui}. 
The vertex truncation is the same as for the $R^4$-case, such that we could compare 
solutions on a box with a known infinite volume/continuum limit. The result is also 
shown in figure \ref{ghostglue-vol}. We obtained ghost and gluon propagators on the 
box which are qualitatively similar to the lattice results. Moreover, by varying the 
volume of the box we found that the volume dependence of these solutions is very
weak, so that the infinite volume/continuum limit is approached extremely slowly. 
Thus it seems as if there is a genuine difference between propagators on different 
manifolds. Since there are indications \cite{Pawlowski:2003hq} that the \lq true' 
exponent $\kappa$ may be closer to $\kappa \approx 0.5$ and therefore closer to the 
current lattice data than our value $\kappa \approx 0.596$, the differences 
shown in figure \ref{ghostglue-vol} may serve as a measure of the upper limit of these 
effects.
 
\begin{figure}[t]
\centerline{
  \epsfig{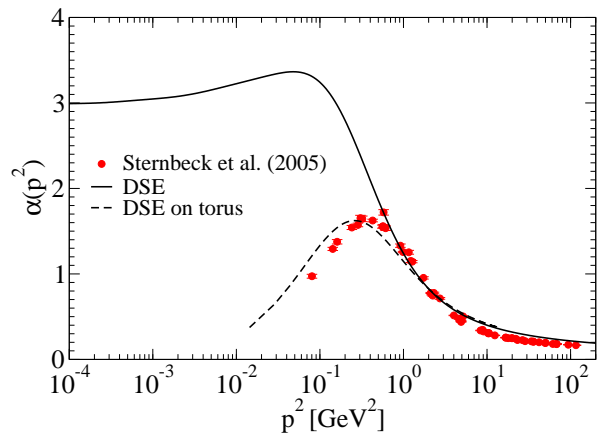}}
\caption{Results for the running coupling of the ghost-gluon vertex from
DSEs in the continuum and on the torus compared to lattice data.
From: Fischer and Alkofer \cite{Fischer:2002hn}) (DSE),
Fischer and Pennington \cite{Fischer:2005nf} (DSE on torus),
Sternbeck {\it et al.} \cite{Sternbeck:2005tk} (lattice).
\label{alpha-vol}} 
\end{figure}

The running coupling resulting from the ghost and gluon dressing functions
of figure \ref{ghostglue-vol} is shown in figure \ref{alpha-vol}. In the continuum
solution we observe a fixed point in the infrared, which corresponds to the analytic
value (\ref{fixed}). The results from DSEs on a compact manifold and the lattice
show an infrared vanishing coupling\footnote{A similar result has been reported for the 
coupling from the three-gluon vertex on the lattice in \cite{Boucaud:2005ce}.}. Although 
the quantitative agreement between the DSE-solution on the torus and the lattice 
data may well be accidental, the qualitative agreement suggests that finite 
volume and boundary effects may play an important role for the infrared behaviour 
of the running coupling.

The running coupling as it results from numerical solutions for the
gluon and ghost propagators can be accurately represented by
\cite{Fischer:2003rp}
\begin{equation}\hspace*{-2.5cm}
\alpha_{\rm fit}(p^2) = \frac{1}{1+(p^2/\Lambda^2_{\tt QCD})}
\left[\alpha_S(0) + (p^2/\Lambda^2_{\tt QCD}) \frac{4 \pi}{\beta_0}
\left(\frac{1}{\ln(p^2/\Lambda^2_{\tt QCD})}
- \frac{1}{p^2/\Lambda_{\tt QCD}^2 -1}\right)\right] \, , \hspace*{3mm} 
\label{fitB}
\end{equation}
with $\beta_0=(11N_c-2N_f)/3$ and the probably unphysical bump in the coupling
at 500 MeV has been omitted. The form of the cancellation of the Landau pole at $p^2=\Lambda^2_{\tt QCD}$
is reminiscent of analytic perturbation theory \cite{Shirkov:1997wi}.  
The expression (\ref{fitB}) is analytic in the complex $p^2$ plane except on the real 
timelike axis where the logarithm produces a cut. This brings us to our next 
topic: the analytical structure of the gluon propagator.

\subsection{The analytical structure of the gluon propagator \label{analytic}}

An important problem in quantum field theory is the question of the separation
of physical and unphysical degrees of freedom. In linear covariant gauges,
where the state space of QCD necessarily is equipped with an indefinite metric
this question is related to the task of specifying a physical, positive
definite subspace $\cal{W}_{\rm phys}$, as discussed in subsection 
\ref{conf}. BRST symmetry and the consequential BRST quartet 
mechanism serve to show that longitudinal gluon and ghost states are orthogonal 
to all states in $\cal{W}_{\rm phys}$ and therefore do not contribute to 
physical $S$-matrix elements. However, this is not automatically guaranteed
for transverse gauge bosons. In fact, transverse, massive gauge bosons are physical
particles in the Higgs phase of Yang-Mills theory. Thus one has to understand
the details of the mechanism by which transverse gluons are taken out of the 
spectrum in the confined phase.  

The question whether the gluon propagator is infrared finite or vanishing is
decisive in this context. This can be seen easily as the relation,
\begin{equation} 
  0 = D(p=0) = \int {d^4x} \; \Delta(x) \,, \label{zero}
\end{equation}
(with $D(p) = Z(p^2)/p^2$) implies that the propagator function in
coordinate space, the Schwinger function $\Delta(x)$, must contain positive as well as 
negative norm contributions, with equal integrated strengths. Thus if the transverse
propagator is vanishing in the infrared, (\ref{zero}) immediately tells us that 
transverse gluons cannot be part of the positive definite, physical state 
space of Yang-Mills theory. The generation of infrared power laws with $\kappa > 0.5$ 
is therefore a viable candidate for a mechanism
of gluon confinement \cite{vonSmekal:1998is,Lerche:2002ep,Zwanziger:2003cf,Alkofer:2003jj}.

\begin{figure}[t]
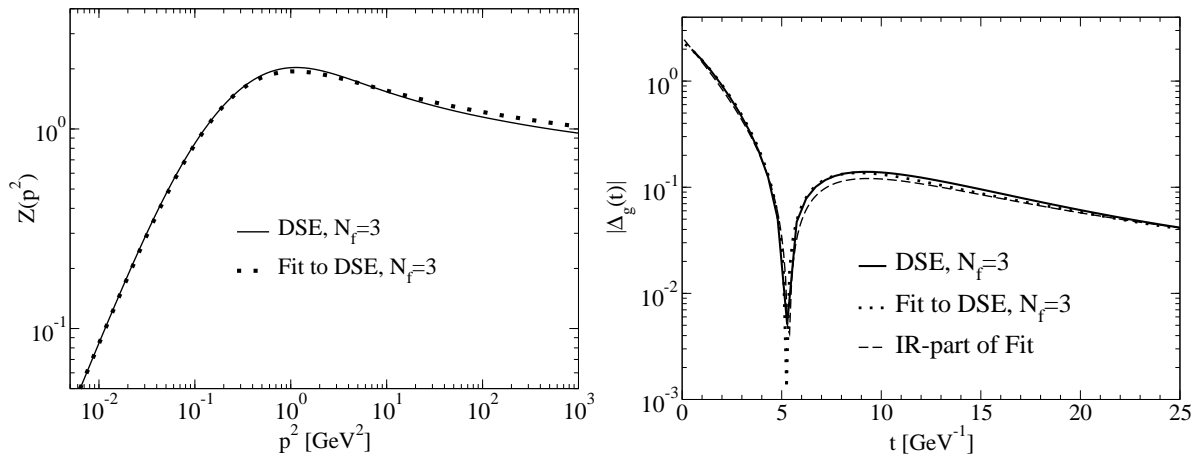

\centerline{\epsfig{file=figure9a.eps,width=7.8cm}
\hfill
\epsfig{file=figure9b.eps,width=7.8cm}}
\caption{Left: The DSE-result for the gluon dressing function
 and the fit (\protect{\ref{fitII}}) are shown. Right: The corresponding
 Schwinger functions (absolute value) of the propagator and the fit.
 From: Alkofer, Detmold, Fischer and Maris \cite{Alkofer:2003jj}.}
\label{xxx}
\end{figure}

The analytic structure of the gluon propagator has been explored
in \cite{Alkofer:2003jj}. The idea is to fit the numerical solutions for the
propagator function $D(p^2)= Z(p^2)/p^2$ (shown in figure \ref{ghostglue}) 
with an analytic expression such that the Fourier transform of the fit also 
reproduces the Fourier transform of the propagator given in (\ref{schwinger}). 
Very good agreement is obtained with the fit form
\begin{equation}
Z_{\rm fit}(p^2) = w \left(\frac{p^2}{\Lambda^2_{\tt QCD}+p^2}\right)^{2 \kappa}
 \left( \alpha_{\rm fit}(p^2) \right)^{-\gamma}\,,
 \label{fitII}
\end{equation}
with $w= 2.65$ and $\Lambda_{\tt QCD}=520$ MeV. The overall magnitude, $w$, depends only
on the renormalisation scale. The ultraviolet anomalous dimension 
$\gamma = (-13 N_c + 4 N_f)/(22 N_c - 4 N_f)$ of the gluon propagator corresponds to
one-loop resummed perturbation theory. The infrared exponent, $\kappa$, is
determined from the infrared analysis, {\it cf.} subsection \ref{IR}. The expression for 
the running coupling, $\alpha_{\rm fit}(p^2)$, has been given in (\ref{fitB}). 
Thus the parameterisation of the gluon propagator has effectively only one parameter, 
the scale $\Lambda_{\tt QCD}$ where the dressing function turns over from the infrared power law
behaviour towards the ultraviolet logarithmic running. The analytic structure of the fit can nicely be
interpreted: The discontinuity across the cut in $Z_{\rm fit}(p^2)$, which vanishes 
for $p^2\to 0^-$, diverges to $+\infty$ at $p^2=-\Lambda_{\tt QCD}^2$ on both sides 
and drops again to zero for $p^2\to -\infty$. This cut can be interpreted as the possible 
decay of the transverse gluon into a ghost-antighost pair or into two or three 
nonperturbative gluons.

Both the propagator and its fit are shown in figure \ref{xxx}
(shown are the unquenched results which are discussed further in
subsection \ref{unquench}). The Schwinger function, $\Delta_g (t)$, based 
on the fit (\ref{fitII}) is compared to the DSE solution in the right diagram
of figure \ref{xxx}. To enable a logarithmic scale, the absolute value is displayed.
In both diagrams the agreement of the numerical solutions with
the fit is excellent. This gives some confidence that at least on a qualitative
level the analytic properties of the Landau gauge gluon propagator are uncovered by 
the fit. The Schwinger function $\Delta_g (t)$ has a zero for 
$t\approx 5 \, {\rm GeV}^{-1} \approx  1 \, {\rm fm}$ and is negative for larger 
Euclidean times. Thus one clearly observes positivity violations in the DSE gluon 
propagator at the expected scale for confinement. This scale is related to the
position of the bump in the gluon propagator and therefore to the position
of the breakdown of the infrared asymptotic theory discussed in subsection \ref{IR}.
This underpins the abovementioned suggestion by Zwanziger that fluctuations at a scale
of $\Lambda_{QCD}$ trigger confinement in Landau gauge. 

\setcounter{footnote}{0}
\section{Dynamical chiral symmetry breaking \label{DXSB}}

Quarks are the central building blocks for mesons and baryons and provide their
primary quantum numbers. In this context chiral symmetry and its breaking pattern
are of great importance in our understanding of the structure and the spectra of
light mesons and baryons. Most of the mass of these objects is generated
dynamically, as we will see in detail later on.

Similar to the situation in the Yang-Mills sector there is an interesting interplay
between the Dyson-Schwinger approach and lattice simulations. Since the strengths
and weaknesses of both approaches are complementary, it is fruitful to combine
both methods to explore the details of chiral symmetry breaking in QCD.

In principle, the lattice is an appropriate nonperturbative tool to study the 
effects of dynamical chiral symmetry breaking. Lattice actions implementing 
overlap, domain wall or perfect fermions obey the Ginsparg-Wilson relation, which
ensures that a lattice variant of chiral symmetry is satisfied. In practice, however,
lattice simulations with reasonably small quark masses are extremely 
expensive in terms of CPU-time. It is only with staggered fermion actions that 
quark masses not too far from their physical values have been achieved to date.
However, these actions have the disadvantage that full chiral symmetry is only 
recovered in the continuum limit and there is no certainty with any finite volume 
that the correct breaking pattern can be observed \cite{Chandrasekharan:2004cn}.

Dyson-Schwinger equations offer a suitable alternative method to investigate
the effects of dynamical chiral symmetry breaking. Since the method is formulated
in continuum field theory no finite volume effects are present in the first place
(see however subsection \ref{torus} for results on a compact manifold). Moreover,
arbitrarily small quark masses can be implemented and the chiral limit can be 
directly investigated without any need for extrapolations. Via the quark propagator
one has direct access to the chiral condensate, the order parameter of dynamical 
chiral symmetry breaking. The truncations in the quark DSE, however, have to be 
controlled. This can be done by comparison with lattice results at quark masses
feasible on the lattice.

In the last section I summarized some results on the structure of Yang-Mills theory,
that have been discovered in the past years. We are now ready to discuss the impact
of these results on the quark sector of QCD. I will first summarize some properties of
the quark propagator in quenched approximation, before I consider the backreaction of the
quarks on the Yang-Mills sector of QCD in subsection \ref{unquench} and compare to
corresponding lattice results. This comparison will be extended in subsection 
\ref{torus}, when I discuss a formulation of the quark DSE on a compact manifold.
As a byproduct of this investigation we will see that a minimal volume for
chiral perturbation theory (and chiral symmetry breaking on the torus in general)
can be obtained from the DSEs. The section ends with a short discussion
of the analytical properties of the quark propagator.

  \subsection{Quark DSE and quark-gluon vertex \label{quark}} 

\begin{figure}[t]
\vspace{0.5cm}
\centerline{
\epsfig{file=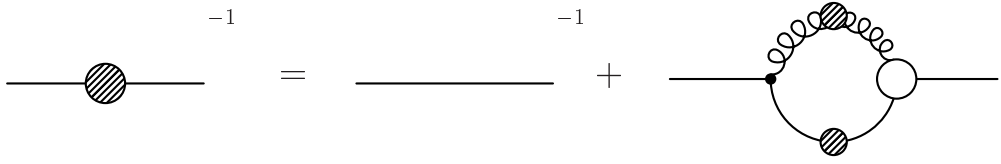,width=13cm}
}
\caption{\label{quark_dse} A diagrammatical representation of the quark Dyson-Schwinger equation.
On the left hand side we find the inverse dressed quark propagator. The diagrams on the right hand
side denote the inverse bare propagator and a dressing loop containing dressed quark and gluon
propagators and one bare and one dressed quark-gluon vertex.
}
\end{figure}

The renormalized Dyson-Schwinger equation for the dressed quark propagator $S(p)$ is given by
\beq
S^{-1}(p) = Z_2 \, S^{-1}_0(p) + g^2\, Z_{1F}\, C_F\, \int \frac{d^4q}{16\pi^4}\,
\gamma_{\mu}\, S(q) \,\Gamma_\nu(q,k) \,D_{\mu \nu}(k) \,,
\label{quark1}
\eeq
with the momentum routing $k=q-p$. Here $D_{\mu \nu}$ denotes the dressed gluon
propagator, $\Gamma_\nu(q,k)$ the dressed quark-gluon vertex and $S_0(p)$ the bare quark
propagator. $Z_2$ and $Z_{1F}$ are renormalisation factors of the quark propagator and 
the quark-gluon vertex. The factor $C_F=(N_c^2-1)/2N_c$ stems from the colour trace of the 
loop. All dependences of the renormalisation point are treated implicitly.
A diagrammatical representation of this equation is given in figure \ref{quark_dse}.

The dressed quark propagator can be written as
\beq
S(p)  = \frac{1}{-i \pslash A(p^2) + B(p^2)} = Z_f(p^2)\frac{i \pslash + M(p^2)}{p^2 + M^2(p^2)} \,,
\eeq
where $A(p^2)$ and $B(p^2)$ are the vector and scalar dressing functions of the quark. The inverse 
$Z_f(p^2) = 1/A(p^2)$ of the vector dressing function is also called the quark wave function 
renormalisation. The ratio 
\beq
M(p^2):=B(p^2)/A(p^2) \label{mass}
\eeq
is the quark mass function. It is important to note that the dependence of $A(p^2)$ and $B(p^2)$ 
on the renormalisation point $\mu^2$ cancels out in the ratio (\ref{mass}). The quark mass function is therefore 
a renormalisation group invariant. The bare quark propagator is given by
\beq
S_0(p) = \frac{1}{-i \pslash + Z_m \: m_R} 
\eeq
and contains the renormalised 'current quark mass' $m_R$ and the mass renormalisation factor $Z_m$.

The two tensor structures in the quark propagator, $\pslash$ and 1, behave different under chiral 
symmetry transformations. The vector part $\pslash$ is invariant, whereas the scalar part is not. 
A nonvanishing function $B(p^2)$ signals unambiguously that chiral symmetry is broken
and therefore may serve as an order parameter. Since $B(p^2)$ is determined by the quark DSE (\ref{quark1}),
this equation is also called the {\it gap equation} of QCD. Another order parameter for dynamical
chiral symmetry breaking is the chiral condensate. It can be shown analytically \cite{Miransky:1985ib}
that the ultraviolet behaviour of the quark propagator in the chiral limit $Z_m \: m_R \rightarrow 0$
can be described by\footnote{The relation of this expression to the Banks-Casher equation 
is discussed by Langfeld {\it et al.} in \cite{Langfeld:2003ye}.}
\beq
M(p^2) = \frac{2 \pi^2 \gamma_m}{3}
\frac{-\langle \bar{\Psi}\Psi\rangle}{p^2 \left(\frac{1}{2} \ln(p^2/\Lambda^2_{QCD})\right)^{1-\gamma_m}} \,.
\label{chiral-M_UV}
\eeq
Here $\gamma_m = \frac{12}{11N_c-2N_f}$ is the anomalous dimension of the quark and 
$\langle \bar{\Psi}\Psi\rangle$ denotes the renormalisation point independent
chiral condensate. This condensate is related to the renormalisation point dependent condensate from
the trace of the quark propagator.
\beq
-\langle \bar{\Psi}\Psi\rangle(\mu^2) := Z_2(s,L) \, Z_m(s,L) \, N_c \, \mbox{tr}_D \int 
\frac{d^4q}{(2\pi)^4} S_{ch}(q^2,s) \,,
\label{ch-cond}
\eeq
by a simple logarithmic factor 
\beq
\langle \bar{\Psi}\Psi\rangle(\mu^2) = \left(\frac{1}{2}\ln(\mu^2/\Lambda^2_{QCD})\right)^{\gamma_m}
\langle \bar{\Psi}\Psi\rangle \,, 
\label{ch-loop}
\eeq
provided the renormalisation point $\mu^2$ is taken large enough. The trace $\mbox{tr}_D$
in (\ref{ch-cond}) is over Dirac indices and $S_{ch}$ denotes the quark propagator in the chiral 
limit. Thus there are two ways to extract the chiral condensate from $S_{ch}$: either by performing 
a fit to the ultraviolet tail of the mass function or from evaluating (\ref{ch-cond}) numerically. 
Both procedures agree with each other. 

The two external ingredients in the quark DSE (\ref{quark1}) are the gluon propagator and the quark-gluon
vertex. While the nonperturbative gluon propagator is a well known object by now
({\it cf} section \ref{YM}), it is the nonperturbative structure of the quark-gluon vertex 
which is at the focus of contemporary studies. This vertex can be decomposed in a 
basis of twelve tensor structures\footnote{There are three independent four vectors: $\gamma_\mu$,
$p_\mu$ and $q_\mu$, and four types of scalars: 1, $\pslash$, $\qslash$, 
$[\pslash, \qslash]$. Combined together these enumerate to twelve 
independent tensor structures \cite{Ball:1980ay}.} 
\beq
\Gamma_\mu = i g \left(\sum_{i=1}^4 \lambda_i L_{\mu}^i
+ \sum_{i=1}^8 \tau_i T_{\mu}^i\right).
\eeq 
Most of the properties of the nonperturbative 
dressing of these structures are not even qualitatively known. Contemporary lattice studies of
the vertex \cite{Skullerud:2003qu} still have large error bars and cannot uncover the 
infrared properties of the vertex. It is well known, however, that sizeable dressing effects
have to be present. A bare quark-gluon vertex is not capable to trigger dynamical chiral 
symmetry breaking. This result underlines the fact that dynamical chiral symmetry breaking 
is an entirely nonperturbative phenomenon. The necessary interaction strength in the vertex 
could be concentrated in the (perturbatively) leading structure  
\beq
L_{\mu}^1 = \gamma_\mu \,,
\eeq
or it could be distributed among several tensor components. One of the potentially important other 
components of the vertex is the scalar piece, 
\beq
L_{\mu}^3 = i(p_1+p_2)_\mu \,,
\eeq
which is proportional to the sum of the incoming and outgoing quark momenta $p_1$ and $p_2$. 
This structure is not invariant under chiral transformations in contrast 
to the leading $\gamma_\mu$-part of the vertex. It only appears when chiral symmetry is broken 
dynamically and provides then a significant self-consistent enhancement of dynamical chiral 
symmetry breaking in the quark DSE \cite{Fischer:2003rp}. 

Basically there have been two different strategies to assess the quantitative impact of
different tensor structures of the vertex on the quarks. First, ansaetze for the vertex 
have been constructed that satisfy approximate forms of the Slavnov-Taylor identity for 
the quark-gluon vertex and avoid kinematical singularities \cite{Fischer:2003rp}. 
Second, a number of attempts have been made to determine parts of the
vertex from solutions to approximate forms of the vertex DSE 
\cite{Bhagwat:2004kj,Bhagwat:2004hn,Fischer:2004ym,Maris:2005tt}. To date the latter approach 
has not yet reached the state of selfconsistency, but seems to be promising for the future.
In this review I focus on the first strategy.

The Slavnov-Taylor identity (STI) for the (colour stripped) ghost-gluon vertex 
$\Gamma_\nu(q,k)$ is given by \cite{Marciano:1978su}
\begin{equation}
G^{-1}(k^2) \: k_\nu \: \Gamma_\nu(q,k) = S^{-1}(p) \: H(q,p) - H(q,p) \: S^{-1}(q),
\label{quark-gluon-STI}
\end{equation}
where $q$ and $p$ are the quark momenta. This identity connects the vertex to the 
quark propagator $S(p)$ and the ghost-quark scattering kernel $H(q,p)$. The presence of the 
ghost dressing function $G(k^2)$ in the STI tells us that the vertex may be an infrared singular
object similar to the three- and four-gluon vertices \cite{Alkofer:2004it} ({\it cf.} subsection
\ref{IR}). An explicit ansatz for the quark-gluon vertex built along this identity has been 
constructed in \cite{Fischer:2003rp}. Its most important parts read 
\beq
\Gamma_\nu(q,k) \sim  G^2(k) \left[\frac{A(p)+A(q)}{2} \gamma_\nu 
+ i \frac{B(p)-B(q)}{p^2-q^2} (p+q)_\nu + (\dots) \right],
\label{vertex_CP}
\eeq
{\it i.e.} it contains a vector as well as a scalar part with relative strength given by the
Abelian approximation to the STI (\ref{quark-gluon-STI}). With this vertex ansatz and 
numerical solutions for the ghost and gluon propagators the quark DSE is closed and can be 
solved numerically.

\begin{figure}[t]
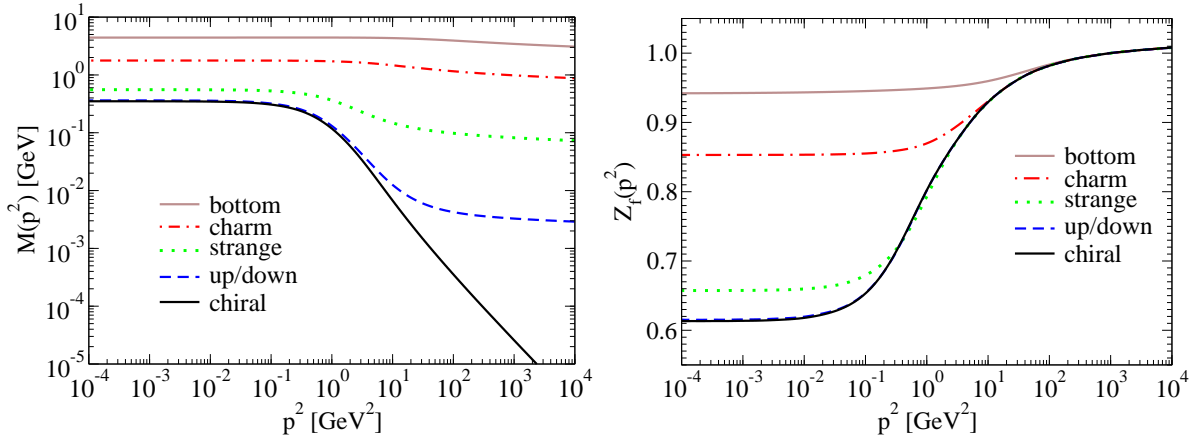

\centerline{
\epsfig{file=figure11a.eps,width=7.8cm}
\hfill
\epsfig{file=figure11b.eps,width=7.8cm}
}
\caption{Results for the quark mass function $M(p^2)$ and the wave function
$Z_f(p^2)$ from the quark DSE for various physical current quark masses.
Based on: Fischer and Alkofer \cite{Fischer:2003rp}.
\label{quarkmass}}
\end{figure}

Results for realistic quark masses can be seen in figure \ref{quarkmass}. 
By comparison of the quark mass function of a light (up/down) quark with the 
chiral limit we see that virtually all of the infrared mass of the up/down quark 
is generated dynamically. These objects are thus dominated by effects due
to dynamical chiral symmetry breaking.  Dynamical effects are still large for the 
strange quark. However, already the charm quark and certainly the bottom quark 
are more or less static objects in the sense that dynamical effects are 
overwhelmed by the large current quark mass and its evolution according to the 
renormalisation group. 

For the quark wave function $Z_f(p^2)$ a large fraction of dynamical mass is signalled
by the size of the dip in the infrared. Whereas the wave function for the up-quark is 
a nontrivial object it becomes more and more static the heavier the quarks are. 
In general we see that the DSE-solutions naturally connect the nonperturbative infrared 
with the perturbative ultraviolet momentum region. What appears as a 'current quark' at 
large momentum transfer and as a 'constituent quark' at small momenta is described by 
the very same object: the dressed quark propagator. 

At large momenta the quark mass functions scale logarithmically with momentum according 
to their expected behaviour from resummed perturbation theory (this can be shown analytically 
in the quark DSE \cite{Miransky:1985ib}). In the chiral limit, where this logarithm is 
absent, one can directly observe the asymptotic behaviour (\ref{ch-cond}) and extract the chiral condensate.
The most recent value for the 
condensate from the DSE-approach (with a lattice based interaction, see subsection \ref{torus})
is given by \cite{Fischer:2005nf}
\beq
-\langle \bar{\Psi}\Psi \rangle^{\overline{MS}} (\mu^2) \;=\; (253.0 \pm 5.0 \mbox{MeV})^3,
\eeq
employing $\Lambda^{\overline{MS}}_{QCD}=0.225(21) \mbox{MeV}$ \cite{Capitani:1998mq}. The
quoted error combines numerical and scale errors but does not include systematic
errors due to the approximation of the quark-gluon vertex. It is therefore interesting to 
compare the central value with recent results from the lattice. Gimenez {\it et al} 
\cite{Gimenez:2005nt} find $(265 \pm 27 \mbox{MeV})^3$ from an operator product expansion 
employing an $O(a)$-improved quenched Wilson action. Wennekers and Wittig \cite{Wennekers:2005wa} 
quote $(285 \pm 9 \mbox{MeV})^3$, determined from a quenched overlap action. Given that 
systematic uncertainties exist in all three approaches one can say that the values are
in fair agreement with each other.

\subsection{Unquenching effects \label{unquench}}

\begin{figure}[t]
\centerline{
\epsfig{file=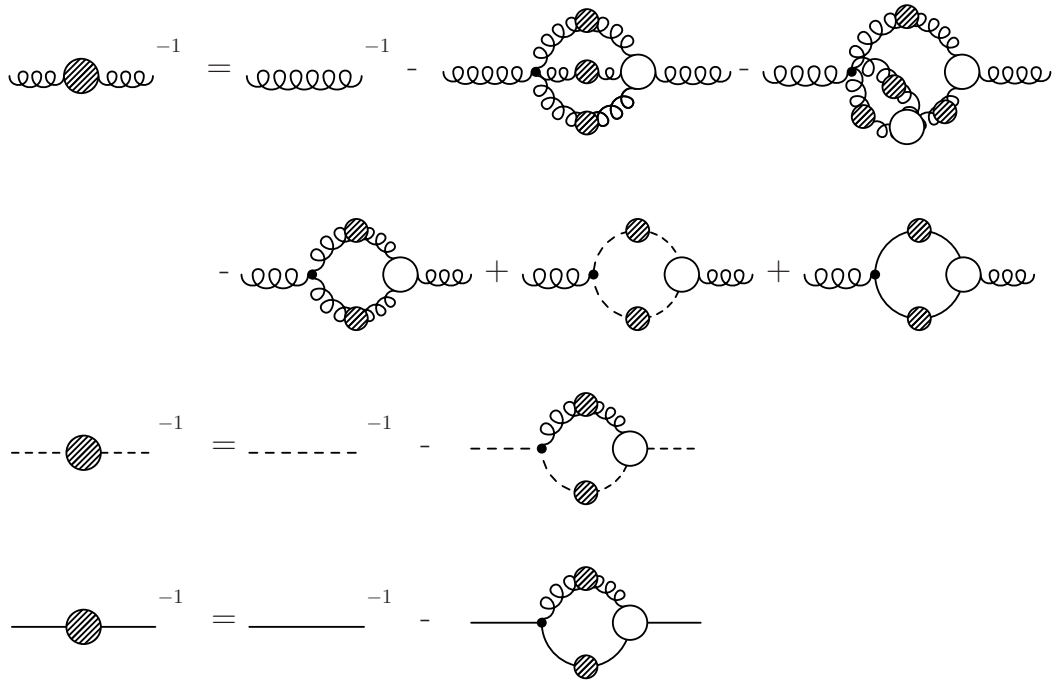,width=14cm}
}
\caption{The coupled set of Dyson-Schwinger equations for the ghost, gluon
and quark propagators. (The tadpole diagram in the gluon DSE has been omitted, since it
drops out in the process of renormalisation.) \label{dse_full}}
\end{figure}

Unquenching is expected to have a significant impact on the properties of QCD if the number of
fermion flavours $N_f$ is large. QCD ceases to be asymptotically free for a sufficiently 
large number of flavours $N_f$. Furthermore a chiral phase transition to the 
symmetric phase is expected. However, the critical $N_f$ are expected to be of 
$\mathcal{O}(10)$\footnote{A recent calculation of Gies and J\"ackel in the framework of the 
exact renormalisation group results in a critical number of flavours of $N_f^c = 10.0\pm 0.4$ 
for the chiral transition \cite{Gies:2005as}.}, which is large compared to the physical case
of three light flavours. Realistic effects due to unquenching may therefore be not too large.

\begin{figure}[t]
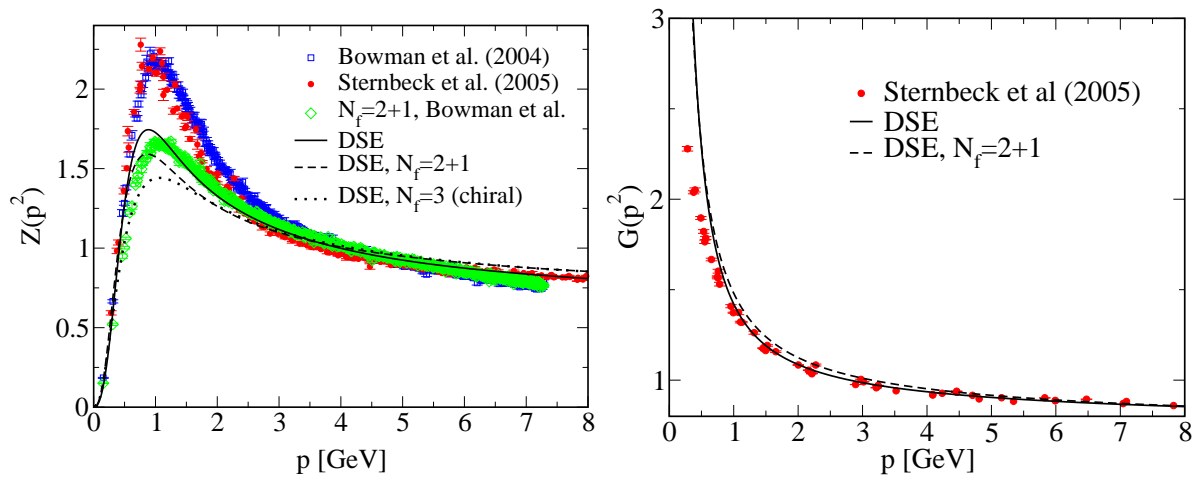

\centerline{
\epsfig{file=figure13a.eps,width=7.8cm}
\hfill
\epsfig{file=figure13b.eps,width=7.8cm}
}
\caption{Unquenched gluon and ghost propagators compared to quenched results.
For the ghost propagator unquenched lattice data are not yet available. 
From: Fischer and Alkofer \cite{Fischer:2003rp} (DSE), 
Bowman {\it et al.} \cite{Bowman:2004jm} (lattice),
Sternbeck {\it et al.} \cite{Sternbeck:2005tk}) (lattice).
\label{unq_YM}}
\end{figure}

For the ghost, gluon and quark propagators of QCD these effects have been determined recently in the DSE-approach 
\cite{Fischer:2003rp,Fischer:2005en}. The effects in the gluon propagator have since been 
confirmed on the lattice \cite{Bowman:2004jm,Bowman:2005vx,Furui:2005mp}; no unquenched
lattice results for the ghost propagator are available yet.
In the Dyson-Schwinger approach one solves the 
coupled set of three equations, which includes a quark loop in the gluon DSE, see 
figure \ref{dse_full} for a diagrammatical representation. Compared to the case of pure
Yang Mills theory, figure \ref{DSE-gl}, there is an additional quark-loop in the 
gluon DSE. This has some impact on the intermediate momentum region as can 
be seen from the numerical results shown in figure \ref{unq_YM}.
In the region around 1 GeV enough energy is present to generate dynamical quark-antiquark 
pairs from the vacuum. These provide some colour screening, which partly eliminates the
antiscreening effects from the gluon self interaction. Consequently the bump in 
the gluon dressing function decreases. This effect is clearly present in both 
the DSE and the lattice study. In the chiral limit the screening effect of the 
quark loop becomes stronger as the energy needed to create a quark pair from the 
vacuum becomes smaller with decreasing bare quark mass. In the ultraviolet momentum 
region unquenching effects are only visible in modified anomalous dimensions as expected 
from resummed perturbation theory.

It is, however, important to note that the inclusion of three light flavours has no effect on the 
infrared structure of the Yang-Mills part of QCD. This can be understood easily in terms
of our infrared analysis from subsection \ref{IR}. Quark loops in the Yang-Mills sector
contain at least two massive quark propagators, which are proportional to $[p^2 + M(p^2)]^{-1}$.
Provided chiral symmetry is broken dynamically we always have sizeable quark masses in these
propagators that dominate at small momenta. Each massive propagator therefore decreases the degree
of infrared singularity of the quark loop by a factor of $p^2$ compared to ghost and gluon loops.
It is hard to see how the quark-gluon vertex should compensate for this. Therefore quark loops
are almost certainly subleading in the infrared in their respective DSEs and cannot influence
the infrared behaviour of Yang-Mills theory\footnote{Certainly this need not be true for the
chirally symmetric phase of QCD.}.

\begin{figure}[t]
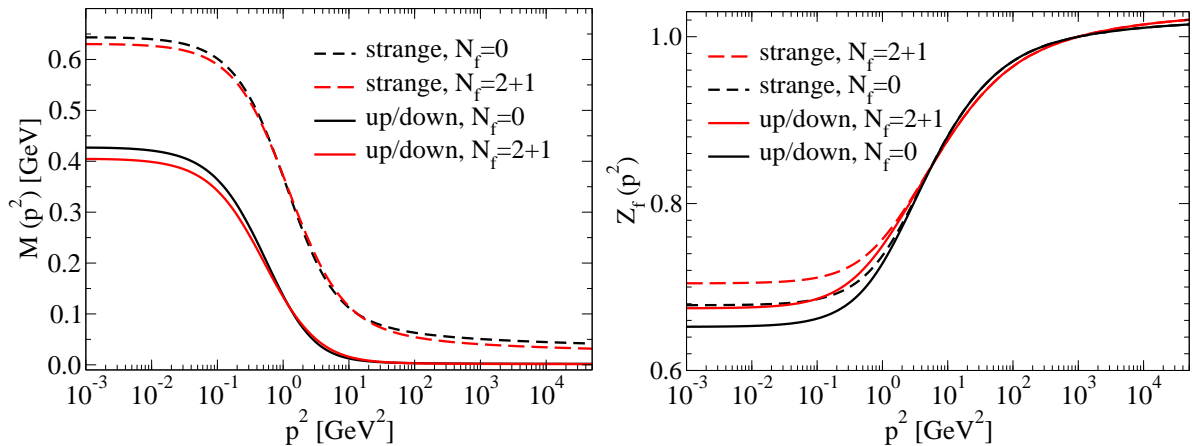

\centerline{
\epsfig{file=figure14a.eps,width=7.8cm}
\hfill
\epsfig{file=figure14b.eps,width=7.8cm}
}
\caption{Unquenched quark propagator compared to quenched results.
From: Fischer, Watson and Cassing \cite{Fischer:2005en}.
\label{quark_unq}}
\end{figure}

In the quark DSE unquenching effects are mediated only indirectly via the quark-gluon 
interaction, {\it i.e.} via the gluon propagator and the quark-gluon vertex. The resulting
effects are much less pronounced than in the Yang-Mills sector, see figure \ref{quark_unq}. 
The quark mass function in the infrared is reduced by roughly $10 \%$ compared to the 
quenched case. This reduction has been seen for a range of truncations for the quark-gluon 
vertex \cite{Fischer:2003rp} and is also confirmed by a lattice study employing staggered 
quarks \cite{Bowman:2005vx}. The chiral condensate from unquenched DSE-solutions is hardly
dependent on the number of flavours as long as $N_f \le 3$ \cite{Fischer:2003rp,Fischer:2005en}. 
Since the condensate is an order parameter for dynamical chiral symmetry breaking it is expected to change rapidly 
at the vicinity of the chiral phase transition. One may therefore conclude that the critical 
number of flavours for the chiral transition at zero temperature is much larger than 
$N_f = 3$. This agrees with our discussion at the beginning of this subsection.

\subsection{Quarks in a box \label{torus}}

Lattice simulations are, of course, always performed at a finite volume. This per se
causes troubles in implementing chiral symmetry. In the introduction to this section
I mentioned the need for lattice actions to satisfy the Ginsparg-Wilson relation.
This ensures that a lattice variant of chiral symmetry is satisfied. However, even
if this relation is satisfied there is more trouble ahead: chiral symmetry is
continuous and can therefore not be spontaneously broken at any finite volume $V$.
This means that a small 'seed' quark mass always has to be present on the lattice.
Chiral symmetry is restored in the limit of zero quark mass, $m \rightarrow 0$, 
independently of the formulation of the lattice action. Thus one has to perform the 
limit $V\rightarrow \infty$ first, before one can investigate the chiral limit 
\cite{Leutwyler:1992yt}. In turn this means that lattice studies employing small quark 
masses are only meaningful if the volume of the underlying manifold is large enough. 
This is one of the reasons why it is extremely expensive in terms of CPU-time to 
simulate small quark masses. Volume effects are therefore a very important issue in 
the investigation of chiral symmetry breaking on compact manifolds.

Chiral perturbation theory on finite volumes is a reliable tool for the extrapolation of
lattice results for meson and baryon observables (see e.g. 
\cite{Detmold:2001jb,Procura:2003ig,Colangelo:2005gd} and references therein). However, 
chiral perturbation theory has nothing to say about volume effects in the underlying 
quark and gluon substructure. Furthermore, chiral perturbation theory builds upon the 
chiral limit, {\it i.e.} it can only be applied on volumes large enough such that small 
quark masses remain accessible. To this end the discrete momentum space induced by the
boundary conditions of the box has to look almost like a continuum one. Correspondingly
the box has to be large enough to allow for sufficiently small nonzero momenta.
On a compact manifold the bosonic degrees of freedom 
have momenta ${\bf p} = 2\pi {\bf n}/L$ with ${\bf n}$ a vector of integers. Small 
nonzero momenta below a typical chiral symmetry breaking scale of $4 \pi f_\pi$ are 
therefore only present if the condition 
\beq
L >> \frac{1}{2 f_\pi} \sim 1 {\rm ~fm}
\eeq
is satisfied. Here $f_\pi$ is the decay constant of the pion. 
{\it A priori} there is no way to say by how much $L$ has to exceed 1 fm
\cite{Colangelo:2005gd}. We will see shortly that this scale can be well estimated 
using the quark DSE on a torus.

\begin{figure}[t]
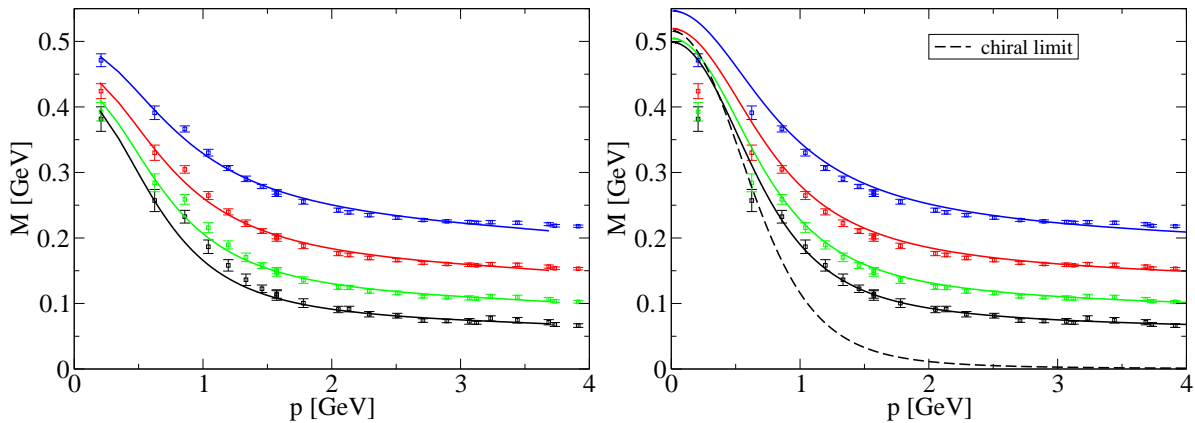

\centerline{
  \epsfig{file=figure15a.eps,width=7.8cm}
  \hfill
  \epsfig{file=figure15b.eps,width=7.8cm}}
\caption{Results for the quark mass function $M$ from Dyson-Schwinger equations 
compared with lattice data for overlap quarks. The DSE-solutions on the left 
diagram are obtained on a similar manifold as the lattice data, whereas in the 
right diagram infinite volume/continuum solutions are shown.
From: Fischer and Pennington, \cite{Fischer:2005nf} (DSE), 
Zhang {\it et al.} \cite{Zhang:2004gv} (lattice).
  \label{fitMZ}}
\end{figure}

In the Dyson-Schwinger approach volume effects can be studied continuously from 
very small to very large volumes by solving the DSEs on a compact manifold. Furthermore
one has direct access to the infinite volume/continuum limit without extrapolations
by solving the very same equations on $R^4$. One is thus in a position 
to study chiral symmetry restoration at small volumes together with effects 
at large and infinite volumes in one common framework. This idea is quite new
and so far volume effects for the ghost and gluon propagator \cite{Fischer:2002eq,Fischer:2005ui} 
({\it cf.} subsection \ref{prop}) and the quark propagator \cite{Fischer:2005nf} 
have been investigated. In the following I shortly summarize the results of the latter study.

As mentioned earlier, the quark-gluon vertex is the key element in the quark DSE.
Following an idea of Bhagwat {\it et al.} \cite{Bhagwat:2003vw} we modelled a simple
approximation to this vertex such that quenched lattice data for various quark masses 
are reproduced on their manifold, {\it i.e.} by solutions to the quark DSE on a torus 
with the same volume as the lattice calculations. The results for the quark mass 
function is shown in the left diagram of figure \ref{fitMZ}, whereas the corresponding 
infinite volume/continuum results are shown in the right diagram of figure \ref{fitMZ}. 
The DSE-solutions on the torus reproduce the lattice data nicely. For small momenta, the 
DSE-results in the infinite volume/continuum limit differ sizeable from the results 
on the compact manifold. It is interesting to see that only the first two lattice points 
for each of the quark masses are affected by the finite volume. All other points
follow the continuum solution. Similar effects have been found for a comparison with
staggered quarks on the lattice. As a result we note that current
lattice simulations may underestimate the amount of dynamical quark mass generation 
in the infrared by as much as 100 MeV.

\begin{figure}[t]
\centerline{
  \epsfig{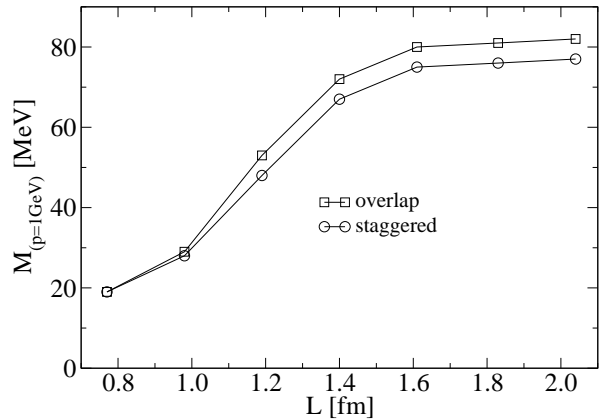}}
\caption{The quark mass function at a given momentum $p=1\mbox{GeV}$ 
plotted as a function of the box length $L$. 
From: Fischer and Pennington, \cite{Fischer:2005nf}. 
\label{L}}
\vspace{3mm}
\end{figure}

A second interesting application of DSEs on a compact manifold concerns the abovementioned estimate
of a minimal box length for chiral perturbation theory. To this end we employed a current 
quark mass of the order of a typical up/down-quark mass and determined the mass function 
$M(p^2)$ at $p^2=1$ GeV from solutions on tori with different volumes. The result is shown
in figure \ref{L}. One observes that the quark mass function grows rapidly in the range 
$1.0 \, <\, L \, <\,  1.6$~fm signalling the onset of dynamical chiral symmetry breaking. 
Above $L\,=\, 1.6$~fm, a plateau is reached. This picture does not change when the mass 
function is extracted at smaller momenta or when even smaller quark masses are employed.  
Thus a safe value for the minimal box length $L$ should be at least  
\beq
L_{\chi PT}\; \simeq\; 1.6 \,\,\mbox{fm}.
\eeq
This provides some justification for extending chiral perturbation theory to rather small 
volumes.

\subsection{The analytical structure of the quark propagator \label{analytic_quark}}

We have seen in subsection \ref{analytic} that the transverse gluon cannot 
be part of the physical asymptotic state space of QCD, because its spectral function contains 
negative norm contributions. It is certainly interesting to investigate whether the same is true for 
the quark propagator. If the answer is yes, this constitutes an independent sign for 
quark confinement, no matter whether a linear rising potential for static quarks can be 
identified in Landau gauge or not ({\it c.f.} the discussion in subsection \ref{conf}). The 
spectral properties of the quark propagator have been studied in a number of publications 
(see {\it e.g.} \cite{Krein:1990sf,Burden:1991gd,Maris:1991cb,Roberts:1994dr,Burden:1997ja} 
and references therein). Here I will concentrate on the most recent results, taken from 
\cite{Alkofer:2003jj,Alkofer:2003jk,Bhagwat:2003vw,Fischer:2005nf}.

The method to determine the analytic structure of the quark propagator is the same as for
the gluon propagator: one calculates the propagator for positive Euclidean momenta and determines
the corresponding Schwinger function (\ref{schwinger}). One can then fit
appropriate forms to the Schwinger function which reflect different analytical structures.
A suitable form for the time dependence of the Schwinger function $\sigma(t)$ of (\ref{schwinger}) is
(see \cite{Alkofer:2003jj} for details) 
\beq
\sigma(t)\; =\; b_0 \exp(-b_1 t) \cos(b_2 t+b_3) \quad , \label{cc}
\eeq
It can be shown that the exponential decay parameter $b_1$ corresponds to the real part of the
leading singularity in the quark propagator. The parameter $b_2$ measures whether
there is an imaginary part. If $b_2=0$ we have a positive definite quark with one pole
on the real momentum axis, whereas $b_2 \ne 0$ corresponds to a pair of complex conjugate
poles in the timelike momentum plane. This second form leads to negative norm contributions 
and therefore describes a confined quark. 

\begin{figure}[t]
\centerline{
  \epsfig{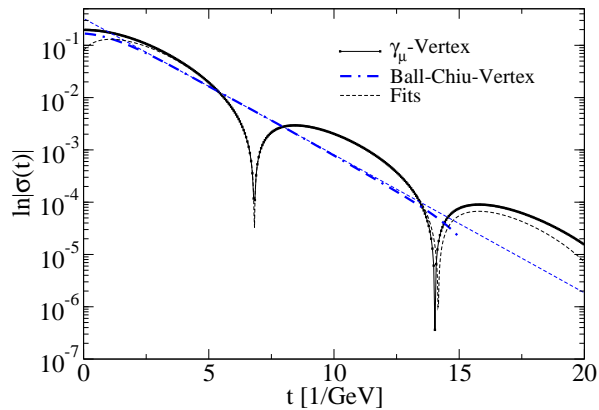}}
\caption{The logarithm of the Schwinger function $\ln(|\sigma(t)|)$ of the chiral
limit quark propagator as a function of time. Shown are results, (i) employing a 
quark gluon vertex with a $\gamma_\mu$-part only and (ii) substituting a construction 
of the form (\ref{vertex_CP}), which contains a strong scalar interaction. 
The results are compared to fits of the function (\ref{cc}). 
From: Fischer and Pennington, \cite{Fischer:2005nf}. \label{schwinger_pic}}
\vspace{3mm}
\end{figure}

Unfortunately it turns out that the quark is a much more difficult case than the gluon. 
The reason is to be found in the structure of the quark-gluon vertex. It has been demonstrated in
\cite{Alkofer:2003jj,Fischer:2005nf} that a sufficiently strong presence of a scalar part 
of the vertex has a significant impact on the analytical structure of the quark 
propagator. This is demonstrated in 
figure \ref{schwinger_pic}. Shown are results for the logarithm of the Schwinger function 
$\sigma(t)$, (i) employing a quark gluon vertex with a $\gamma_\mu$-part only and (ii) 
substituting a construction of the form (\ref{vertex_CP}), which contains a strong scalar 
interaction as well. The oscillations seen for case (i) correspond to a quark propagator 
with complex conjugate poles. These are located at a \lq quark mass' of 
$m = 516(20) \pm i \, 428(20)$ MeV. The spikes in the curve indicate sign changes in the
Schwinger function signalling negative norm contributions and therefore quark confinement.
On the other hand, the exponentially decaying Schwinger function for case (ii) corresponds 
to a positive definite quark propagator with a pole on the real axis at 
$m = 632(20) \pm i \, 0(2)$ MeV (within numerical accuracy). Although such a quark may
still be confined via the BRST-quartet mechanism \cite{Nakanishi:1990qm}, this cannot be seen
from the analytical structure of the quark propagator.

These results show that the analytic structure of the quark propagator depends
strongly on the details of the structure of the quark-gluon vertex.
A crucial point seems to be whether the scalar parts in the dressed quark-gluon 
vertex are of similar strength as in the ansatz (\ref{vertex_CP}) or not. Lattice data for
the quark propagator alone cannot decide this question, since these can be reproduced by both
a vertex with or without a strong scalar contribution \cite{Fischer:2005nf}. It seems as 
if the only way to solve this problem is to analyse the quark-gluon vertex directly either on the lattice 
or by solving its DSE. First attempts in this direction on the lattice \cite{Skullerud:2003qu}
as well as in the DSE approach \cite{Bhagwat:2004hn,Bhagwat:2004kj,Fischer:2004ym}
have been made, but have not yet reached a stage which allows definite statements.
The analytical structure of the quark propagator therefore remains yet an open problem.

\setcounter{footnote}{0}  
\section{Light Mesons as bound states of quarks and gluons \label{mesons}}

In the last two sections I summarized the results of recent efforts to determine some
fundamental properties of the elementary particles of QCD, the quarks and gluons. 
We have seen that their nonperturbative two-point functions are significantly
different from their counterparts in perturbation theory. Certainly this comes as 
no surprise. Strong quantum fluctuations are to be expected to affect the 
properties of quarks and gluons since these are confined. In this section 
we will consider the impact of these results on observable properties of light 
mesons. The framework emerging from these efforts directly connects structures from 
the Yang-Mills- and the quark-sector of QCD with the experiment.
 
I will first focus on probably the most important aspect of QCD when it comes to 
light mesons: the emergence of pseudoscalar mesons as the Goldstone bosons
of dynamically broken chiral symmetry. I will then summarize recent efforts to
explicitly include the gluonic substructure into the description of these mesons.
This allows for a first investigation of unquenching effects in the pion and 
the rho meson. Finally I will summarize a broad range of results on  
experimental observables that have been obtained in the last years employing a 
simple model for the quark-gluon interaction. This last subsection is intended 
only as a short guide to the wealth of existing literature on this subject. 

\subsection{Goldstone bosons and quark-antiquark states \label{goldstone}}
  
The pion as a bound state of a quark and an antiquark is described by the 
homogeneous Bethe-Salpeter equation (BSE) which can be written schematically
as
\beq
\Gamma^\pi(p;P)\;=\;\int \frac{d^4k}{(2 \pi)^4}\, 
K(p,k;P)\, S(k_+)\, \Gamma^\pi(k;P)\, S(k_-).
\label{eq:bse}
\eeq
Here $\Gamma^\pi(p;P)$ is the Bethe-Salpeter amplitude of the pion and $K(p,k;P)$ 
is the so called Bethe-Salpeter kernel, which describes the interaction of the
quark and the antiquark, with relative momentum $p$, inside a pion of momentum $P$.
The momentum arguments  $k_+=k+\xi P$ and $k_-=k+(\xi-1)P$ of the two quark 
propagators are defined such that the total momentum of the pion is given by 
$P=k_+-k_-$. All physical results are independent of the momentum partitioning 
$\xi=[0,1]$ between the quark and the antiquark.
Equation (\ref{eq:bse}) can be easily adapted to describe mesons of arbitrary
flavour content. The Bethe-Salpeter amplitude of any pseudoscalar particle
including the pion can be decomposed into four Dirac-structures,
\beq
\hspace*{-5mm}
\Gamma^{PS}(p;P) \;=\; \gamma_{5} \left(\Gamma_0(p;P) - \imath \Pslash \Gamma_1(p;P)
-\imath \pslash \Gamma_2(p;P) - \left[\Pslash,\pslash \right]\Gamma_3(p;P) \right)\,,
\eeq
which can be determined separately from the BSE (\ref{eq:bse}) once the 
interaction kernel $K$ has been specified.

The crucial link between the meson bound states and their quark and gluon constituents 
is provided by the axial vector Ward-Takahashi identity (axWTI). Denoting the quark DSE 
({\it cf.} figure \ref{quark_dse}) by
\beq
S^{-1}(p)= S_0^{-1}(p) - \Sigma(p)
\eeq
one can write the axial vector Ward-Takahashi identity as
\beq
-\left[\Sigma(p_+)\gamma_5+\gamma_5\Sigma(p_-)\right] =
\int\frac{d^4k}{(2\pi)^4} K(p,k;P)
\left[\gamma_5 S(k_-) + S(k_+) \gamma_5 \right],
\label{eq:axwti}
\eeq
where again all flavour and spinor indices have been omitted. We see that this 
identity demands a tight relation between the quark self-energy $\Sigma$ and 
the Bethe-Salpeter kernel $K$. Maris, Roberts and Tandy have shown analytically
that this relation ensures that the pion is a massless Goldstone boson in the 
chiral limit \cite{Maris:1997hd,Maris:1998hd}. Subsequently Bicudo {\it et al}
\cite{Bicudo:2001jq} and Bicudo \cite{Bicudo:2003fp} established an analytic proof that 
Weinberg's low energy theorems for $\pi-\pi$ scattering, the Goldberger Treiman 
relation and the Adler zero in the chiral limit hold in all approximation schemes 
for the quark DSE and the Bethe-Salpeter kernel $K$ that satisfy the axWTI. This 
establishes a profound understanding of the chiral properties of the pion in
terms of the underlying gauge theory.  

One example of a truncation scheme satisfying the axWTI is the well studied 
rainbow-ladder truncation (for reviews see \cite{Maris:2003vk,Maris:2005tt}). 
This truncation employs a specific form of the quark gluon vertex, which 
leads to rainbow-like diagrams in the quark DSE together with
ladder-like diagrams in the Bethe-Salpeter equation. The Goldstone-boson
nature of the pion is also manifest in the numerical solutions to such 
truncation schemes, as we will see in the next subsection.

\subsection{Unquenching light mesons \label{unquenching}}

\begin{figure}[t]
\vspace{0.5cm}
\centerline{
\epsfig{file=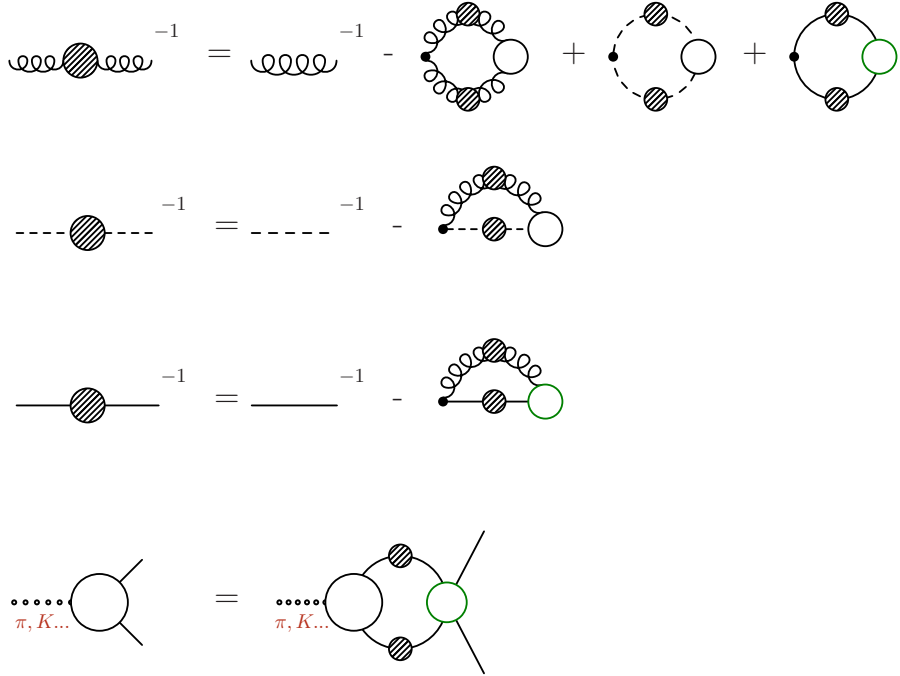,width=12cm}
}
\caption{\label{SDEs} A diagrammatical representation of the coupled system of 
ghost, gluon and quark Schwinger-Dyson equations and the meson Bethe-Salpeter
equation.  Filled circles denote dressed propagators and empty circles denote 
dressed vertex functions.
}
\end{figure}

As an example of an application of the DSE/BSE formalism I discuss 
recent attempts to implement and quantify unquenching effects in the description 
of light mesons. Depending on the observable under investigation these effects
are qualitatively and quantitatively important. Unquenching is mandatory to
describe meson decays like $\rho \rightarrow  \pi \pi$. They are also
anticipated to be important for scalar mesons where one deals with genuine
quark-antiquark states or (admixtures of) meson-meson or even diquark-diquark
correlations \cite{Watson:2004jq,Pennington:2006qi}. The latter contributions are only present in 
the unquenched theory.

\begin{figure}[t!]
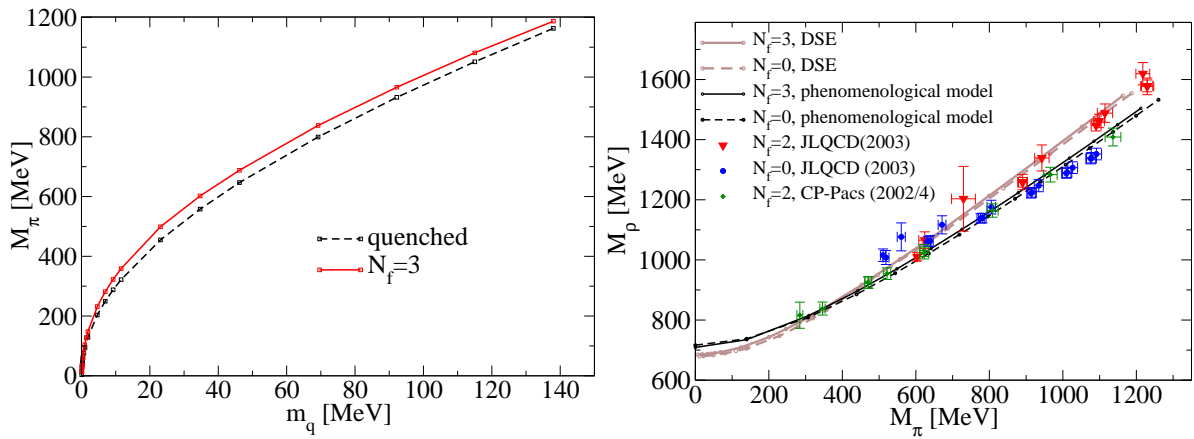

\centerline{
\epsfig{file=figure19a.eps,width=7.8cm}
\hfill
\epsfig{file=figure19b.eps,width=7.8cm}}
\caption{\label{fig:plot1} Left diagram: pseudoscalar meson masses as functions 
of the current quark mass for the quenched and unquenched theory with three 
degenerate sea quarks. Right diagram: Quenched and unquenched results for the
vector meson mass as a function of the pseudoscalar meson mass.
From: Fischer, Watson and Cassing \cite{Fischer:2005en} (DSE/BSE), 
CP-PACS collaboration \cite{AliKhan:2001tx,Namekawa:2004bi} and 
JLQCD collaboration \cite{Aoki:2002uc} (lattice).}
\end{figure}
\begin{table}[b]
\caption{\label{tab:fpar}Parameter sets and results for the masses $m_{\pi}$, 
$m_K$, $m_{\rho}$ and decay constants $f_{\pi}$, $f_K$
for the quenched case ($N_f=0$), the unquenched case 
with three degenerate 'sea'-quarks ($N_f=3$) and the physical quark configuration 
case ($N_f=2+1$) with two up/down quarks and one strange quark. The values for 
the current quark masses are evolved to $\mu = 2\,\mathrm{GeV}$ according to 
their one-loop behaviour. All units are given in MeV.
From: Fischer, Watson and Cassing \cite{Fischer:2005en}.}
\begin{indented}
\lineup
\item[]\begin{tabular}{@{}*{8}{l}}
\br                              
&$m_u$ & $m_s$ & $m_{\pi}$ & $f_{\pi}$ & $m_K$ & $f_K$ & $m_{\rho}$ \cr
\mr
$N_f=0$   &4.17  &88.2   &139.7 &130.9 &494.5 &165.6 &708.0         \cr
$N_f=3  $ &4.06  &       &139.7 &130.8 &      &      &690.0         \cr
$N_f=2+1$ &4.06  &86.0   &140.0 &131.1 &493.3 &169.5 &695.2         \cr
PDG\cite{Eidelman:2004wy}&&    &139.6 &130.7 &493.7 &160.0 &770.0     \cr
\br                              
\end{tabular}\end{indented}\end{table}

A systematic examination of unquenching effects in the quark-quark four-point 
function (related to the Bethe-Salpeter kernel $K$ of equation (\ref{eq:bse})) 
has been carried out by Watson and Cassing \cite{Watson:2004jq}. They demonstrated 
that the unquenched four-point function contains physical resonances only, whereas 
unphysical diquark states are absent. Subsequently a quantitative analysis of a 
particular class of unquenching effects (quark loops in the gluon polarisation) 
has been carried out in \cite{Fischer:2005en}. To this end the coupled system of 
DSEs and BSEs shown diagrammatically in figure \ref{SDEs} has been solved. The 
employed quark-gluon vertex is a simplified version of (\ref{vertex_CP}). 
This vertex and the corresponding Bethe-Salpeter kernel are chosen such that the 
axWTI (\ref{eq:axwti}) is satisfied. The resulting pion is a massless Goldstone
boson in the chiral limit as can be seen in the left 
diagram of figure \ref{fig:plot1}. Plotted is the pion mass as a function of the
current quark mass at a fixed large renormalisation point. In the chiral
limit, {\it i.e.} for $m_q \rightarrow 0$, one obtains $M_\pi \rightarrow 0$
without any finetuning.

A suitable quantity to compare the results with lattice data is the mass of the 
$\rho$-meson as a function of the pion mass. No scheme ambiguities arise since 
both quantities are physical. The results are shown in figure \ref{fig:plot1}. 
Compared are results from the full DSE-setup, described above, with those 
obtained employing a phenomenological model for the quark-gluon interaction (the 
details of the model are described in \cite{Fischer:2005en}). The model interaction 
and the quark-gluon interaction of the complete set of DSEs are complementary to 
each other in the sense that the model interaction is confined to a quite narrow 
momentum region, whereas the interaction of the full DSE-setup has considerable 
strength in the infrared and extends into the ultraviolet according to the 
perturbative one-loop scaling. Together, both setups represent a measure for 
the theoretical error of the calculation. This error is of the same size as the 
combined systematic error of the different lattice simulations. In general, the 
results are in nice agreement with the lattice data. Below 240 MeV lattice data
are not yet available. The results from the DSEs show a nonlinear dependence of 
the vector meson mass on the pseudoscalar one. The effect of unquenching -- when 
viewed as a function of the pseudoscalar meson mass -- becomes the same for both 
schemes: the vector meson mass is slightly increased when quark loops are taken 
into account. This trend is also seen in the lattice simulations, where the effect 
is even more pronounced.  However, these unquenching effects are small compared 
to the differences between both the two truncation schemes that have been 
employed and the systematic errors of the lattice results. 

The comparison of the results from the DSE/BSE approach with the lattice is 
encouraging, though there is still a lot of work to be done. The current truncation
scheme for the DSE/BSEs has to be extended systematically to include the full tensor
structure of the quark-gluon interaction. Furthermore the unquenched quark four-point
function of \cite{Watson:2004jq} has to be incorporated to allow for effects due
to the decay of the $\rho$-meson. Together, both improvements can be expected to bring
the $\rho$-mass to its experimental value at the physical point in figure 
\ref{fig:plot1}. The resulting $\rho-\pi$-mass curve should then agree with the one 
from chiral perturbation theory at small pion masses and with (potentially improved)
lattice data at large pion-masses. If so, one has a framework which allows to
extract and understand the internal structure of light mesons in terms of
(nonperturbative) quarks and gluons.

\subsection{Meson properties from BSEs \label{rainbow}}

The framework described in the last subsection explicitly resolves the details of 
quark and gluon propagation and their interaction inside mesons. Such an approach 
is feasible at the expense of considerable technical and numerical effort. A technically
simpler approach is to approximate the combination of the gluon propagator and the 
quark-gluon vertex by a model function. This approach provides
interactions which are within the class of rainbow/ladder approximations that
satisfy the axWTI. Thus the correct behaviour of the pion as a Goldstone boson is 
guaranteed. These models have been explored extensively in the past years, leading 
to a range of interesting results. Some of these have been discussed in detail in recent 
reviews \cite{Maris:2003vk,Maris:2005tt}. I will only give a very brief overview 
here, which is intended as a short (and by far not exhaustive) guide to the literature.

The employed models for the combined gluon propagator and quark-gluon vertex
are built upon the known perturbative ultraviolet limit and possess a certain
interaction strength in the infrared which is controlled by a number of
parameters (typically two or three). The central premise of such an account is 
the idea that the detailed shape of the interaction in the infrared is not 
important in the quark DSE. The integrated strength of this effective interaction 
between the quarks has to be sufficiently strong to induce spontaneous chiral 
symmetry breaking and associated dynamical quark mass generation. This idea has 
well known limitations in the scalar and axialvector meson sector 
\cite{Bender:1996bb,Alkofer:2002bp,Watson:2004kd} but turned out to be successful 
for the description of pseudoscalar and vector mesons.

Explicit calculations within such an approximation scheme explained the twofold 
nature of the pion as Goldstone boson and bound state of heavy constituents 
\cite{Maris:1997hd,Maris:1998hd}. The simultaneous absence of diquark states from 
the physical spectrum of (Landau gauge) QCD has been demonstrated in 
\cite{Bender:1996bb}\footnote{The corresponding absence of bound diquark states in 
Coulomb gauge has been reported recently in \cite{Alkofer:2005ug}.}. Masses and decay 
constants of vector mesons have been determined in \cite{Maris:1999nt}. Corresponding 
results for strange and charmed pseudoscalar and vector mesons are reported in 
\cite{Krassnigg:2004if}. Detailed studies of the electromagnetic properties of 
ground state pseudoscalar mesons (charge radii, form factors, etc.) can be found 
in \cite{Maris:1998hc,Maris:1999bh,Maris:2000sk}. Results for radially excited 
pseudoscalar mesons have been reported recently in \cite{Holl:2005vu}. 

In addition to static properties of mesons also scattering processes have been analysed.
Aspects of $\pi-\pi$ scattering have been investigated in 
\cite{Bicudo:2001jq,Cotanch:2002vj,Bicudo:2003fp} ({\it cf.} subsection \ref{goldstone}). 
In particular the existence of the $\sigma$ and $\rho$-resonances at the proper energies 
in $\pi-\pi$ scattering has been shown in \cite{Cotanch:2002vj}. In references 
\cite{Maris:1998hc,Kekez:1998rw} it has been demonstrated that the cross section of the 
anomalous decay $\pi \rightarrow \gamma \gamma$ known from the perturbative triangle 
diagram is reproduced in rainbow/ladder approximation. Processes including 
the  $\gamma-3\pi$ form factor are investigated 
in \cite{Alkofer:1995jx,Bistrovic:1999yy,Bistrovic:1999dy,Cotanch:2003xv}.
The results agree well with corresponding low energy theorems. Explicit decays like 
$\rho \rightarrow \pi\pi$, $\phi \rightarrow KK$ and $K^* \rightarrow \pi K$ are 
investigated in \cite{Jarecke:2002xd}. Aspects of the $U_A(1)$-problem and
$\eta-\eta'$-mixing are analyzed in \cite{vonSmekal:1997dq,Klabucar:1997zi,Kekez:2000aw}. 
Conceptual issues and results for pion quark distributions have been 
discussed in \cite{Hecht:2000xa,Tiburzi:2003ja}.

\section{Concluding remarks}

In this review I summarized recent results obtained from the framework of
Dyson-Schwinger and Bethe-Salpeter equations in Landau gauge QCD. One of the issues
I emphasized is the capability of this method to build bridges between
different areas of quantum field theory. The approach connects the perturbative 
ultraviolet momentum regime with the nonperturbative low energy limit of the theory. 
It also connects hadron phenomenology with an underlying description of hadrons in 
terms of dressed quarks and gluons. The approach is complementary in its strengths 
and weaknesses to lattice gauge theory and therefore provides an alternative tool 
to analyse the theoretical structures of QCD.

The results in the Yang-Mills sector of QCD support a possible infrared effective 
theory which is given by the gauge-fixing parts of the action only. Ghost degrees 
of freedom dominate in the infrared and provide for long range correlations whereas 
the gluon propagator vanishes at zero momentum. We have seen that the latter property
implies positivity violations in the gluon propagator. Transverse gluons are
therefore confined. We analysed the behaviour of the running coupling of 
SU(N)-Yang-Mills theory in the infrared and found a fixed point which is 
(qualitatively) universal and invariant at least in a class of transverse gauges.

In the quark sector we discussed the effects of dynamical chiral symmetry breaking 
in the dressed quark propagator. Most of the mass of light quarks is generated 
dynamically. The chiral condensate can be extracted reliably from the quark propagator.
Chiral symmetry breaking on a compact manifold has been studied. Probably the most 
notable result here is a minimal box length for chiral perturbation theory. 
In the meson sector of QCD we saw that the approach naturally reproduces
the Goldstone nature of the pion as well as resulting low energy theorems. The 
applicability of the framework as a tool for hadron phenomenology is well explored.
First steps have been made to explicitly investigate effects from the gauge sector
of QCD in meson observables.

There are also a number of open problems that pose challenges for the future.
The gluon self-interaction may be a key ingredient in our understanding
of the transition from the perturbative to the nonperturbative region of
Yang-Mills theory and needs to be further investigated. The detailed structure
of the quark-gluon interaction and its consequences for the analytical structure
of the quark propagator is not yet clarified. When it comes to comparison with 
lattice results an important open problem is the difference between the continuum 
results for the ghost and gluon propagators and the results on the compact manifold. 
Here one likes to see how exactly the continuum limit is approached when the volume 
of the compact manifold is increased. In general, it also seems promising to further explore 
the relation of the results in Landau gauge to other gauges as for example Coulomb 
gauge or the maximal Abelian gauge. Finally, unquenching the meson sector is a 
central issue in order to establish even closer contact with experiment.

\ack
It is a pleasure to thank Alan Martin for the invitation to write this review. 
I am grateful to Reinhard Alkofer, Wolfgang Cassing, Will Detmold, Holger Gies, 
Felipe Llanes-Estrada, Pieter Maris, Mike Pennington, Hugo Reinhardt, Lorenz von Smekal, 
Peter Watson and Dan Zwanziger for pleasant and fruitful collaborations 
on some of the topics discussed here. Furthermore I would like to thank 
Mandar Bhagwat, Jacques Bloch, John Gracey, Andreas Krassnigg, Derek Leinweber, 
Orlando Oliveira, Jan Pawlowski, Craig Roberts, Paulo Silva, Jonivar Skullerud, 
Peter Tandy and Tony Williams for inspiring discussions and
Reinhard Alkofer and Dominik Stoeckinger for a critical 
reading of the manuscript.   
This work has been supported by the GSI, Darmstadt and the Deutsche 
Forschungsgemeinschaft (DFG) under contract Fi 970/7-1.


\section*{References}
\bibliographystyle{utcaps.bst}
\bibliography{revbib}

\end{document}